\documentclass[12pt,a4paper]{article}

\usepackage[T1]{fontenc} 
\usepackage{ae,aecompl}  
\usepackage{amsmath}
\usepackage{amssymb}
\usepackage{accents}
\usepackage{booktabs}  
\usepackage{longtable} 
\usepackage{lscape}    
\usepackage{xspace} 
\usepackage[usenames]{color} 
\usepackage[flushmargin,hang]{footmisc} 
\usepackage{cite} 
\usepackage{array}
\usepackage{a4}
\usepackage{a4wide}
\usepackage{epsfig}
\usepackage{caption}
\usepackage{subfig}
\usepackage{psfrag} 
\usepackage{dcolumn} 
\usepackage[noconfig]{refstyle} 
\usepackage{listings}
\usepackage{courier}

\usepackage{fancyhdr}
\newref{eqn}{%
name = Eqn.~,
names = Eqns.~,
Name = Eqn.~,
Names = Eqns.~,
refcmd = \eqref{#1},
rngtxt = --, 
lsttxt = {\space and~}}
\newref{tab}{%
name = Tbl.~,
names = Tbls.~ ,
Name = Tbl.~,
Names = Tbls.~,
rngtxt = --,
lsttxt = {\space and~}}
\newref{sec}{%
name = Sec.~,
names = Secs.~,
Name = Sec.~,
Names = Secs.~,
rngtxt = --, 
lsttxt = {\space and~}}
\newref{bib}{%
key = ,
name = Ref.~,
names = Refs.~,
Name = Ref.~,
Names = Refs.~,
refcmd = \cite{#1},
rngtxt = --, 
lsttxt = {\space and~}}

\usepackage{bbm}
\usepackage{epsfig}
\usepackage{array}
\usepackage{float}
\usepackage{dsfont}
\usepackage{amstext}
\usepackage{rotating}

\psfrag{meff}{{\scriptsize $\langle m_\nu \rangle$ [eV]}}
\psfrag{s12}{{\scriptsize $\sin^2 \theta_{12}$}}
\psfrag{T12}{{\scriptsize $T_{1/2}^{ \, 0 \nu } $ [10$^{27}$ y]}}
\psfrag{dma}{{\scriptsize $\Delta m_{\rm A}^2$ [10$^{-3}$ eV$^2$]}}
\psfrag{NME}{{$M^{\prime 0\nu}$}}
\psfrag{mihmax}{{\scriptsize $\langle m_\nu \rangle_{\rm max}^{\rm IH}$}}

\definecolor{MyDarkGreen}{rgb}{0,0.63,0.23} 
\definecolor{MyDarkBlue}{rgb}{0.23,0.21,0.69} 
\definecolor{MyLightBlue}{rgb}{0.22,0.51,0.86}

\usepackage[colorlinks=true,linkcolor=MyDarkBlue,citecolor=MyDarkGreen,
urlcolor=MyLightBlue,bookmarksnumbered=true,bookmarksopen,breaklinks]{hyperref} 
\hypersetup{bookmarksopenlevel=0}

\newdimen\LTcapwidth \LTcapwidth=\textwidth
\newcommand{\nubb}{$0\nu\beta\beta$\xspace}

\def\be{\begin{equation}}
\def\beq{\begin{equation}}
\def\ee{\end{equation}}
\newcommand{\bal}{\begin{align}}
\newcommand{\eali}{\end{align}}
\newcommand{\ba}{\begin{array}{c}}
\newcommand{\bad}{\begin{array}{ccc}}
\newcommand{\ea}{\end{array}}
\newcommand{\bea}{\begin{eqnarray}}
\newcommand{\eea}{\end{eqnarray}}
\newcommand{\meff}{\ensuremath{\langle m_\nu \rangle}\xspace}

\newcommand{\mhiggs}{\ensuremath{m_\mathrm{H}}\xspace}

\newcommand{\dckm}{\ensuremath{\delta_\mathrm{CKM}}\xspace}

\newcommand{\soten}{$SO(10)$\xspace}
\newcommand{\strep}{16\xspace}
\newcommand{\mgut}{\ensuremath{M_\text{GUT}}\xspace}
\newcommand{\msusy}{\ensuremath{M_\mathrm{SUSY}}\xspace}
\newcommand{\mz}{\ensuremath{M_Z}\xspace}
\newcommand{\RaiseBrace}[1]{\raise1.0pt\hbox{$\displaystyle#1$}}
\def\SU{\mathrm{SU}}
\def\U{\mathrm{U}}
\DeclareMathOperator{\Lagr}{\mathcal{L}}

\DeclareMathOperator{\Tr}{Tr}
\def\chisq{\ensuremath{\chi^2}\xspace}
\def\gsim{\mathrel{
   \rlap{\raise 0.511ex \hbox{$>$}}{\lower 0.511ex \hbox{$\sim$}}}}
\def\lsim{\mathrel{
   \rlap{\raise 0.511ex \hbox{$<$}}{\lower 0.511ex \hbox{$\sim$}}}}
\def\barroman#1{\sbox0{#1}\dimen0=\dimexpr\wd0+1pt\relax
  \makebox[\dimen0]{\rlap{\vrule width\dimen0 height 0.06ex depth 0.06ex}%
    \rlap{\vrule width\dimen0 height\dimexpr\ht0+0.03ex\relax 
            depth\dimexpr-\ht0+0.09ex\relax}%
    \kern.5pt#1\kern.5pt}}

\def\be{\begin{equation}}
\def\ee{\end{equation}}
\def\gs{\mathrel{
   \rlap{\raise 0.511ex \hbox{$>$}}{\lower 0.511ex \hbox{$\sim$}}}}
\def\ls{\mathrel{
   \rlap{\raise 0.511ex \hbox{$<$}}{\lower 0.511ex \hbox{$\sim$}}}}

\newcommand{\D}{\displaystyle}

\begin{document}

\title{\vspace{-2cm}
\hfill {\small }\\[-0.1in]
\vskip 0.8cm
\bf \Large
Fits to $SO(10)$ Grand Unified Models
}
\author{
~~Alexander Dueck\thanks{email: 
\tt alexander.dueck@mpi-hd.mpg.de}\mbox{ },~~
Werner Rodejohann\thanks{email: 
\tt werner.rodejohann@mpi-hd.mpg.de}\mbox{
}
\\\\
{\normalsize \it Max-Planck-Institut f\"ur Kernphysik,}\\
{\normalsize \it  Postfach 103980, D-69029 Heidelberg, Germany}  \\
} 
\date{}
\maketitle
\thispagestyle{empty}
\vspace{0.18cm}
\begin{abstract}
\noindent  
We perform numerical fits of Grand Unified Models based on $SO(10)$, using various combinations of 
10-, 120- and 126-dimensional Higgs representations. 
Both the supersymmetric and non-supersymmetric versions are fitted, 
as well as both possible neutrino mass orderings. 
In contrast to most previous works, 
we perform the fits at the weak scale, i.e.~we use RG evolution from the GUT scale, at which 
the GUT-relations between the various Yukawa coupling matrices hold, down 
to the weak scale. 
In addition, the 
right-handed neutrinos of the seesaw mechanism are integrated out one by one in the RG running. 
Other new features are the inclusion of recent results on the reactor neutrino mixing angle and 
the Higgs mass (in the non-SUSY case). As expected from vacuum stability considerations, 
the low Higgs mass and the large top-quark Yukawa coupling cause 
some pressure on the fits. A lower top-quark mass, as sometimes argued to be 
the result of a more consistent extraction from experimental results, 
 can relieve this pressure and improve the fits.   
We give predictions for neutrino masses, including the effective one for 
neutrinoless double beta decay, as well as the atmospheric neutrino mixing angle and the leptonic 
$CP $ phase for neutrino oscillations.

\end{abstract}

\newpage





\tableofcontents

\newpage

\section{Introduction}
\label{sec:introsoten}
The origin of fermion masses and mixings is a long-standing question in elementary particle physics. Among different frameworks to address this problem, theories unifying strong and electroweak interactions as well as -- partly or completely -- quarks and leptons 
offer very attractive solutions. Particularly intriguing are models based on \soten symmetry.

In \soten  all Standard Model (SM) particles of one generation plus a right-handed neutrino are assigned to a single \strep-dimensional representation. The masses of the fermions arise from Yukawa interactions of two 16s with suitable Higgs fields when the latter develop vacuum expectation values (VEVs). Since \cite{Slansky:1981yr,Ross:1985ai}
\be
16 \otimes 16 = 10_\text{S} \oplus \overline{126}_\text{S} \oplus 120_\text{A},
\ee
the Higgs fields of renormalizable \soten models can belong to 10-, 126-, or 120-dimensional representations (denoted henceforth by $10_H$, $\overline{126}_H$, and $120_H$, respectively) yielding the Yukawa part of the Lagrangian, 
\be
\Lagr_Y = 16 \, (Y_{10} 10_H + Y_{126} \overline{126}_H + Y_{120} 120_H ) \, 16 \,.
\ee

After spontaneous symmetry breaking the fermions obtain the masses \cite{Ross:1985ai,Dutta:2004zh,Dutta:2005ni,Senjanovic:2006nc}
\beq
\label{eqn:mrel}
\begin{aligned}
M_u &= v_{10}^u Y_{10} + v_{126}^u Y_{126} + v_{120}^u Y_{120} \, ,\\
M_d &= v_{10}^d Y_{10} + v_{126}^d Y_{126} + v_{120}^d Y_{120} \, ,\\
M_{D} &= v_{10}^u Y_{10} - 3 v_{126}^u Y_{126} + v_{120}^{D} Y_{120} \, ,\\
M_l &= v_{10}^d Y_{10} - 3 v_{126}^d Y_{126} + v_{120}^l Y_{120} \, ,\\
M_R &= v_{126}^R Y_{126} \, ,\\
M_L &= v_{126}^L Y_{126} \, ,
\end{aligned}
\ee
where $M_u, M_d, M_D, M_l, M_R, M_L$ are the up-quark, down-quark, Dirac neutrino, charged lepton, 
right-handed Majorana neutrino (type I seesaw), and left-handed Majorana neutrino (type II seesaw) 
mass matrices. Further, $Y_{10}$, $Y_{126}$, and $Y_{120}$ are the Yukawa coupling matrices between the fermions and the 10-, $\overline{126}$-, and 120-dimensional representations, respectively. Here  
$Y_{10}$ and $Y_{126}$ are symmetric, while $Y_{120}$ is anti-symmetric. The 
$v^{u/d/D/l/R/L}_{10/126/120}$ represent the parts of the VEV (or combination of VEVs) of the Higgs 
fields that are important for the respective particle type. 
Eqn.~\eqref{eqn:mrel} holds at the scale of \soten symmetry breakdown, 
henceforth referred to as $\mgut=2\times10^{16}$ GeV. Fermion masses and mixings are measured at much lower energies, e.g., at $M_Z \approx 91.19$~GeV. Hence, to fit the parameters of a model to the data, one has to use renormalization group evolution (RGE) to obtain the model predictions at \mz.


As obvious from \eqnref{mrel}, in \soten all fermion mass matrices are related since they are combinations of the same Yukawa matrices. A numerical fit, taking into account these relations, and comparing to the 
experimentally determined observables, can test different \soten models for phenomenological viability. 
In this paper we will perform such tests. 
In this regard we consider several classes of \soten models, differing in the choice of Higgs representations. A minimal model with only one Higgs field is phenomenologically not viable since all fermions are then proportional to the same Yukawa matrix and are hence diagonal in the same basis, resulting in no mixing between up- and down-quarks or between charged leptons and neutrinos. A $\overline{126}_H$ field is required for neutrino mass generation via a seesaw mechanism. 
We consider models that additionally contain either a $10_H$, or a $120_H$, or both a $10_H$ and a $120_H$. More model-dependent effects of intermediate scales are neglected in our study. 
We analyze supersymmetric versions of the models as well as models without SUSY.

Fits of \soten models along the lines presented in this paper have of course 
been performed before \cite{Matsuda:1999yx,Dev:2012xn,Das:2000uk,Xing:2007fb,
Ross:2007az,Grimus:2006bb,Fukuyama:2007ri,Altarelli:2010at,Joshipura:2011nn,Buccella:2012kc,Altarelli:2013aqa}. 
Our approach is different from previous works in the following aspects: 

\begin{itemize}
\item[1)]
Firstly, we perform a full renormalization group evolution and fit the models to the experimental 
data at $\mu = \mz$. This means that we start from \eqnref{mrel}, evolve the observables {\it down} 
to \mz, and compare at that scale with available experimental data. 
In contrast, previous studies either did not include renormalization effects at 
all \cite{Matsuda:1999yx,Dev:2012xn}, or evolved experimental values {\it up} 
to \mgut \cite{Ross:2007az,Grimus:2006bb,Fukuyama:2007ri,
Altarelli:2010at,Joshipura:2011nn} and fitted that data at $\mu = M_\text{GUT}$. 
The latter procedure introduces the following issue: to evolve observables from \mz to \mgut, certain 
high-energy model details (such as Yukawa couplings) have to be known, since they have an 
impact on the running of observables (this has been demonstrated long ago, e.g., 
for the $m_b/m_\tau$ ratio \cite{Vissani:1994fy}). However, exactly these model 
details are varied while the fit is performed. Our approach is therefore more consistent.

\item[2)]Secondly, when performing our fits we take into account effects coming from non-degenerate right-handed neutrinos, $\nu_{R_i}$, with mass $M_i$ -- an issue which is commonly neglected in fits to GUTs. 
When performing the RGE one has to integrate out the heavy neutrinos 
at appropriate energies. Since $\nu_{R_i}$ can be highly hierarchical one has to integrate them out one by one at $\mu = M_{i}(M_{i})$ (as opposed to integrating out all at once at a common seesaw scale). This produces several effective field theories (EFT) during RGE -- one EFT per heavy degree of freedom which is integrated out -- with different running behavior of the parameters. 
Treating non-degenerate $\nu_{R_i}$ correctly can have sizable effects on neutrino parameters as has been demonstrated in Ref.~\cite{Antusch:2002rr}. 
For our analysis we apply the method described in Ref.~\cite{Antusch:2002rr} and integrate out $\nu_{R_i}$ at appropriate energies. 
Besides yielding more trustworthy results, a more precise analysis that takes into account RGE also leads to more reliable predictions of experimentally undetermined observables like the effective neutrino mass governing neutrinoless double beta decay or the leptonic $CP$ violating phase $\delta_{CP}^l$. 
To show the impact of our more rigorous treatment of RGEs, we will fit the models also when we 
evolve the experimental values up to \mgut,  fit those values at that scale, use the 
low energy neutrino parameters and ignore the effects of non-degenerate right-handed 
neutrinos (denoted in what follows as "no RGE''). 

\item[3)] The inclusion of RGE allows us to consider the Higgs mass in the analysis of non-SUSY models, since it is related to the other observables via the RG equations 
(described in detail in \secref{higgsmass}). As well-known, and expected, 
from vacuum stability bounds, 
requiring that the Higgs quartic coupling remains positive puts pressure on the fits due to 
the large value of the top-quark mass. We demonstrate the effects on the fits by using a lower value for the top-quark mass and by leaving out the Higgs mass, respectively. The fits are shown to improve considerably. 
\item[4)] Finally, more precise data in the neutrino sector is now available, most notably through 
the discovery of a non-zero reactor mixing angle \cite{Abe:2011fz,An:2012eh,Ahn:2012nd}. 
\end{itemize}
Our fits will assume dominance of the type I seesaw, and will be performed for both possible neutrino 
mass orderings, as well as for the non-supersymmetric and supersymmetric cases.

The paper is build up as follows: In Section \ref{sec:models} we will describe the different 
models that we fit, and in Section \ref{sec:fittingprocedure} give some details on 
the fit procedure and the observables that need to be reproduced. 
Section \ref{sec:results} describes the results 
of the fits, before we conclude in Section \ref{sec:sotenconcl}. The appendices contain for completeness 
the best-fit parameters of the Yukawa matrices, and the $\beta$-functions necessary for the 
RG evolution.  

\section{Model Details and Previous Work}
\label{sec:models}
We now simplify the notation and rewrite \eqnref{mrel} as \cite{Dutta:2004zh,Dutta:2005ni,Altarelli:2010at,Joshipura:2011nn}:
\beq
\label{eqn:yrel}
\begin{aligned}
Y_u &= r ( H + s F + i t_u \, G) \, , \\
Y_d &= H + F + i G  \, ,\\
Y_{D} &= r ( H - 3 s F + i t_D \, G) \, , \\
Y_l &= H - 3 F + i t_l \, G  \, , \\
M_R &= r_R^{-1} F \, ,
\end{aligned}
\ee
where the Yukawa matrices $Y_i$ are the mass matrices $M_i$ from \eqnref{mrel} divided by the VEVs $v$ or $v_{u/d}$ of the Standard Model (SM) or Minimal Supersymmetric Standard Model (MSSM), respectively. $H$, $F$, $G$ correspond to  $Y_{10}$, $Y_{126}$, and $Y_{120}$, respectively, i.e., $H$ and $F$ are symmetric and $G$ is anti-symmetric. The parameters $s, t_u, t_D, t_l$ are complex, whereas $r, r_R$ can be chosen real without loss of generality \cite{Joshipura:2011nn}. We have omitted the type II seesaw term (compare to \eqnref{mrel}) since we assume for definiteness that the type I seesaw term dominates. 
We will consider supersymmetric (SUSY) as well as non-supersymmetric \soten models. In non-SUSY \soten the $10_H$ can be chosen real, but one can argue that this will not lead to a viable particle spectrum \cite{Bajc:2005zf}. Taking the $10_H$ to be complex, its real and imaginary parts can couple separately to the fermionic 16 and will lead to two independent Yukawa matrices. To avoid this complication in the case of non-SUSY \soten, we impose an additional Peccei--Quinn $U(1)$ symmetry \cite{Peccei:1977hh} as described in Refs.\ \cite{Babu:1992ia,Bajc:2005zf,Joshipura:2011nn}. In this case \eqnref{yrel} is valid both in the case of non-SUSY as well as SUSY \soten. 

We now define the different models which we want to test for viability using experimental data on fermion masses and mixings. The first differentiation between the models concerns their Higgs content. We consider two minimal setups with $10_H + \overline{126}_H$ or $120_H + \overline{126}_H$ and a setup with $10_H + \overline{126}_H + 120_H$. We refer to the $10_H + \overline{126}_H$ setup as "MN" ("MS") and to the $10_H + \overline{126}_H + 120_H$ setup as "FN" ("FS") in case we consider the non-SUSY (SUSY) versions of the models (M stands for "minimal", F for "full", 
N for ``non-SUSY'', S for ``SUSY'').  
The non-SUSY $120_H + \overline{126}_H$ setup was argued, based on an analytical two generation approximation, to be an attractive minimal model to describe fermion masses and mixings \cite{Bajc:2005zf}. 
The three generation non-SUSY setup case was later shown to be not successful \cite{Joshipura:2011nn}. 
We analyze this setup numerically within our approach, and confirm that it cannot reproduce 
the observed data, see below. The SUSY case is also analyzed by us, to the best of our 
knowledge for the first time, and shown not to be a valid model either.  
More details on the models are given in the following subsections, the main features are given in 
\tabref{models}.\\

Our analysis involving RGEs that integrate 
out the individual heavy Majorana masses is quite CPU-intensive. For this reason 
we are forced, in the present paper, to ignore some complications arising in \soten models:  

The Higgs representations mentioned above are not enough to break \soten down to the SM. Therefore further Higgs representations are necessary (see also Ref.\ \cite{Altarelli:2013aqa,Mambrini:2013iaa} for recent analyses). In case of non-SUSY models a minimal choice would be to add one $45_H$ \cite{Bertolini:2009qj} and in case of SUSY models one $210_H$ \cite{Goh:2004fy,Melfo:2010gf}. Furthermore, in SUSY models one needs additionally a $126_H$ which keeps SUSY from being broken by D terms \cite{Melfo:2010gf}. 
Since our analysis including the RGEs is already quite involved, we ignore 
details of different viable breaking chains, which would induce new scales, RGEs, parameters, etc. 
Effectively we therefore assume that intermediate symmetries are close to \mgut and the running of parameters between these scales and \mgut is not sizable. Hence, the relevant information for our analysis is the Higgs content given in \tabref{models} and Eqns.~\eqref{eqn:mrel} or \eqref{eqn:yrel}, together with the beta-functions of the SM or MSSM as given in the appendix.

Gauge coupling unification also depends on the breaking chain and the values of intermediate scales. E.g., in the minimal non-SUSY model based on $10_H + \overline{126}_H$, it has been shown 
that with Higgs VEVs of $45_H$ and $\overline{126}_H$ around $10^{14}$~GeV 
\cite{Bertolini:2009es,Bertolini:2009qj} (i.e.\ suitable to reproduce the neutrino mass-squared 
differences via seesaw) gauge coupling unification can be achieved \cite{Bertolini:2012im,Bertolini:2013vta}. 
In contrast, for the SUSY version of the minimal model it has been shown \cite{Bertolini:2006pe} that reproducing known values of neutrino mass-squared differences leads to light states which spoil gauge coupling unification. 
Since we do not analyze the details of \soten breaking, 
we will also not be concerned with the unification of gauge couplings. 
In our fit the gauge couplings (whose 1-loop RGE do only depend on themselves)   
are chosen at the GUT scale such that they reproduce 
their measured values at $M_Z$. 

The results that we will obtain in this paper are therefore all under the assumption of 
negligible effects coming from intermediate scales $M_I$. Those are typically of order $10^{10}$ 
to $10^{11}$ GeV, and the running from the GUT scale to $M_I$ involves 6 orders of magnitude, while 
from $M_I$ to $M_Z$ involves 8 orders of magnitude. Moreover, the gauge couplings influence 
the RGEs, and ignoring their unification will have an effect there as well. 
However, as emphasized in Ref.\ \cite{Altarelli:2013aqa}, the contribution of
the Higgs states with masses around $M_I$ is only a sub-leading correction in the running from 
$M_{\rm GUT}$ to $M_I$, 
because the corresponding beta-function coefficients are small. 
Nevertheless, there are corrections to be expected, but their impact is 
hard to estimate, and would have to be made case-by-case. 
As an example, we can compare with Ref.\ \cite{Altarelli:2013aqa}, in which a fit that takes  
into account intermediate scales and gauge coupling unification 
is performed. That scale is the one at which 
$SO(10)$ is broken to the Pati-Salam group $SU(4)_C \times SU(2)_L \times SU(2)_R$. 
As the values of the fitted observables 
are the same current values that we are using, this analysis is the one we should compare with. 
In our language, the fit in \cite{Altarelli:2013aqa} 
is ``no RGE'' within model MN. With four free parameters less than we have 
(the charged lepton masses and $r_R$ are not fitted), the $\chi^2$-minimum is 
17.4, compared to 1.1 in our case. Large part of the difference of the $\chi^2$-minima can be attributed to the requirement that the baryon asymmetry as generated by thermal leptogenesis is included in the fit of Ref.\ \cite{Altarelli:2013aqa}. Performing the 
fit without the baryon asymmetry indeed gives $\chi^2_{\rm min} \simeq 1$, in very good agreement with our result. As in our case, an inverted hierarchy is 
not possible. Regarding predictions, the atmospheric neutrino mixing parameter 
$\sin^2 \theta_{23}$ is 0.353 in \cite{Altarelli:2013aqa}, and 0.406 here. Both approaches 
predict it to be somewhat low, and cause some pressure on the fits. 
With $r_R$ not fitted in \cite{Altarelli:2013aqa}, the 
predictions for the neutrino masses are not really comparable, but agree within  
factors of a few. It seems that, at least in this particular example, 
the main features of the fits are stable with respect to the 
intermediate scales. There can apparently be shifts of the $\chi^2$, but no dramatic 
shifts that cause a particular model to be ruled out with intermediate scales while being allowed 
without. We should stress however that we cannot guarante this for all models under study.

In what follows we will describe the properties of the models that we fit, summarizing shortly 
the results.

\begin{table}[t]
\centering
\begin{tabular}{@{}lccc@{}} 
\toprule
Higgs content		&	SUSY	&  non-SUSY		& free parameters	\\
\midrule
one of $10_H$, $120_H$, or $\overline{126}_H$	&	\multicolumn{2}{c}{no mixing} & -- \\
\addlinespace
$10_H + 120_H$	&	\multicolumn{2}{c}{no type I seesaw} & -- 	\\ \addlinespace
$120_H+\overline{126}_H$ &  \multicolumn{2}{c}{not valid}  & 	17		\\ \addlinespace
$10_H+\overline{126}_H$ &  MS  & MN	& 19	\\ \addlinespace
$10_H + 120_H + \overline{126}_H$ &  FS  & FN	&	18	\\ 
\bottomrule
\end{tabular}
\caption{Brief overview of considered models, names given to them in the text and the number of free parameters. Models with only one Higgs representation cannot produce mixing, and models without $\overline{126}_H$ do not have a seesaw mechanism. The number of observables that we fit is 18 or 19, depending on whether we include the Higgs mass or not. Models with $120_H+\overline{126}_H$ Higgs 
representations do not provide acceptable fits. }
\label{tab:models}
\end{table}

\newpage
\subsection{Minimal Model with $10_H + \overline{126}_H$ (MN, MS)}
\label{sec:minmod1}
In this model we do not have a $120_H$, hence $G=0$ in \eqnref{yrel}. To count the number of free parameters we choose a basis in which $H$ is real and diagonal, which leaves us with 19 real parameters: 3 in $H$, 12 in $F$ (complex symmetric), and 4 in $r$ (real), $s$ (complex), and $r_R$ (real) (assuming type I seesaw dominance).

There is a plethora of literature about the supersymmetric version of this model 
\cite{Babu:1992ia,Lavoura:1993vz,Brahmachari:1997cq,Bajc:2001fe,Bajc:2002iw,Goh:2003sy,Goh:2003hf,
Aulakh:2003kg,Aulakh:2004hm,Bertolini:2004eq,
Bajc:2004xe,Bajc:2004fj,Aulakh:2005mw,Babu:2005ia,Bertolini:2005qb,
Bajc:2005qe,Bertolini:2006pe,Aulakh:2007jm,Bajc:2008dc,Joshipura:2011nn}, 
which is often referred to as the "Minimal Supersymmetric Grand Unified Theory" (MSGUT)\footnote{An alternative approach containing two 10-dimensional 
Higgs representations can be found in \cite{BhupalDev:2011gi}.}. 
Literature analyzing the non-supersymmetric version also exists \cite{Matsuda:2001bg,Bajc:2005zf,Bertolini:2009qj,
Bertolini:2009es,Joshipura:2011nn,Buccella:2012kc}. The predictivity of this model has been 
pointed out first in \cite{Babu:1992ia}. 
All authors analyzing the fermion spectrum neglect details of the RGE which affect the running of observables between \mz and \mgut, as described in \secref{introsoten}.

{\bf Results:}
In case of an inverted neutrino mass hierarchy both the non-SUSY as well as the SUSY versions of this model are not able to reproduce the data.
In case of a normal hierarchy without including RGE, model MN (non-SUSY) gives a good fit to fermion masses and mixing angles. 
Including RGE and fitting in addition the Higgs mass leads to tension and a somewhat 
unsatisfactory fit. The top-quark mass is $3.4~\sigma$ too small at its best-fit point. 
This is related to its influence on the Higgs quartic coupling, whose positivity till the GUT  
scale requires a smaller top-quark mass than measured. If the top-quark mass was lower, as sometimes argued 
to be the result of a more consistent extraction from the data 
\cite{Langenfeld:2009wd,Hoang:2011TRSlides,Hoang:2008xm}, the fit improves.  
In addition, the leptonic mixing angles $\theta^l_{23}$ and $\theta^l_{13}$ are not reproduced very well. 
The SUSY model MS can fit the data and fits including RGE are even somewhat better than fits without RGE.
Further details will be discussed in \secref{resmin}.

\subsection{Alternative Minimal Model with $120_H + \overline{126}_H$}
Due to absence of $10_H$ we have $H = 0$ in this model. Going to a basis with real diagonal $F$ (3 parameters), we have 6 real parameters in $G$ and 8 in $s, r_R$ (real), $t_u, t_l, t_D$ (complex), altogether 17 parameters (neglecting type II seesaw).

This model is analytically analyzed in Ref.\ \cite{Bajc:2005zf} in the case of only two fermion generations (second and third) and argued to be viable and predictive. A numerical three generation analysis finds the model to be unable to fit fermion masses and mixings \cite{Joshipura:2011nn}. To provide further evidence for this result we perform a fit of this model. In addition to the normal neutrino mass hierarchy considered in Ref.\ \cite{Joshipura:2011nn}, we also try to fit the inverted hierarchy. Further, we include full RGE into our analysis, which has not been done in previous studies. Moreover, we also attempt to 
fit the SUSY version of the model, which to the best of our knowledge has not been done before. 

{\bf Results:}
We confirm with our more consistent fit approach 
that this class of models is not compatible with data on fermion masses and mixing angles, irrespective of the 
neutrino hierarchy or whether they are supersymmetric or non-supersymmetric.

\subsection{Model with "full" Higgs Content $10_H + \overline{126}_H + 120_H$ (FN, FS)}
Without additional constraints, we would have the maximal number of parameters in this model. One can however considerably reduce the number of parameters by assuming all parameters to be real. This can be motivated or derived from an underlying parity symmetry \cite{Dutta:2004hp} or spontaneous $CP$ violation \cite{Grimus:2006rk}. If $CP$ is violated spontaneously solely by purely imaginary VEVs of the $120_H$, this corresponds to taking all parameters in \eqnref{yrel} to be real. We will use the model with this reduced number of parameters and refer to it as "FN" in the non-supersymmetric case and as "FS" in the supersymmetric case. In a basis with real diagonal $H$ (3 parameters) we count 6 parameters in $F$, 3 in $G$, and 6 in $r, s, t_u, t_l, t_D, r_R$, altogether 18 parameters. So in spite of introducing $120_H$ in addition to $10_H$ and $\overline{126}_H$, through the additional constraints this model has one parameter less than the ``minimal'' one. Therefore some authors refer to the SUSY version of this model as the ``New Minimal Supersymmetric GUT'' (NMSGUT) \cite{Aulakh:2006hs}.

As in the minimal model there is a large amount of literature coping with the ability of the SUSY version of this model to reproduce the fermion spectrum and mixing \cite{Aulakh:1982sw,Clark:1982ai,Aulakh:2000sn,Oshimo:2002xg,Oshimo:2003cm,Bertolini:2004eq,
Dutta:2004hp,Bertolini:2005qb,Grimus:2006bb,Grimus:2006rk,
Aulakh:2006hs,Aulakh:2007ir,Aulakh:2008sn,Altarelli:2010at,Joshipura:2011nn,Joshipura:2011rr}. Without invoking supersymmetry, this model is analyzed only in Refs.\ \cite{Joshipura:2011nn,Dutta:2013bvf}.

{\bf Results:}
This class of models gives good fits to the data for both the normal and inverted neutrino mass hierarchy. For the fits of the non-SUSY version of this model the Higgs mass still leads to the top-quark mass being too small at its best-fit point, in this case $3.3~\sigma$ with normal neutrino mass hierarchy and $3.5~\sigma$ with inverted hierarchy, in analogy to the situation mentioned for model MN in Sec.\ \ref{sec:minmod1}. Again, for a smaller top-quark mass the 
fits improve. If the fits include our rigorous treatment of RGEs, they worsen for the non-SUSY case, and to 
a lesser extent also in the SUSY case.

\section{Details of the Fitting Procedure}
\label{sec:fittingprocedure}
We fit the models to experimental values of the masses of quarks and charged leptons, mass-squared differences of neutrinos, and mixing angles of quarks (including $\dckm$) and leptons. The quark and charged lepton masses at $M_Z$ are taken from \bibref{Xing:2007fb}. Since the masses of charged leptons are measured with a very high accuracy that 
goes beyond our 1-loop RGE analysis, and since furthermore such precise values make a numerical fit very challenging, we assume an uncertainty of 5\,\% for these observables when fitting the models to the data. For the neutrino observables we neglect the running below $M_Z$ and take the values from \bibref{GonzalezGarcia:2012sz}\footnote{See also \bibref{Fogli:2012ua,Tortola:2012te}.}. 

To check our numerical algorithm we also make fits without RGE as in
Ref.~\cite{Joshipura:2011nn}. 
That means, as explained in Sec.\ \ref{sec:introsoten},  we ignore the effect of
non-degenerate right-handed neutrinos and 
take experimental values of observables at $\mu =$~\mgut as given in 
\tabref{expvalsnorge, expvalsnorgesusy}, and fit the GUT-relations to those numbers.  
Note that to simplify the fitting procedure we symmetrized the error bars around the best-fit value whenever they were not symmetric. This will not have a large effect on the fits, since strongly non-symmetric errors are present only for the light quark masses where the uncertainty is large anyway. Finally, we perform separate fits for both a normal hierarchy (NH) and an inverted hierarchy (IH) of the neutrino masses (see \secref{nhvsih}). We collect the values of observables underlying our analysis in \tabref{expvals, expvalsnorge, expvalsnorgesusy}. To fit the model parameters to the observables we minimize
\beq \label{eqn:chisqdef}
\chi^2 = \sum_{i=1}^n\left(\frac{y_i^{\rm theo}(x) - y_i^{\rm exp}}{\sigma_i^{\rm exp}}\right)^2 
\ee
numerically with respect to $x = (x_1,...,x_m)$, where $y_i^{\rm exp}$ are observables measured experimentally with uncertainty $\sigma_i^{\rm exp}$, and $y_i^{\rm theo}(x)$ is the corresponding theoretical prediction given the vector $x$ of model parameters. 
We will later also look at \chisq as a function of the atmospheric mixing angle, $\sin^2 \theta_{23}^l$. To derive such a function for an observable $O$, 
one can add a term $(O^{\rm theo}(x) - O)^2/(0.01 \, O)^2$ to the definition of \chisq, where the denominator is a very small artificial uncertainty to let the minimization algorithm converge to a minimum. If $O$ itself was part of the definition of \chisq in \eqnref{chisqdef}, then its term with the experimental uncertainty is removed. After performing the minimization of the so defined $\chisq$-function, one evaluates with the parameters obtained from that fit \chisq as given in \eqnref{chisqdef}, i.e.~without the contribution of the artificial error, but including the contribution of the real experimental uncertainty. This method was previously used in Refs.\ \cite{Grimus:2006rk,Bertolini:2006pe,Joshipura:2011nn}.
%
%
\begin{table}[t]
\centering
{
\renewcommand{\arraystretch}{1.022}
\begin{tabular}{@{}cr@{ }c@{ }l@{}}
\toprule
Observable  & \multicolumn{3}{c}{Exp. value} \\
\midrule
$m_d$ [GeV] & 0.0029 & $\pm$ & 0.001215 \\
$m_s$ [GeV] & 0.055 & $\pm$ & 0.0155 \\
$m_b$ [GeV] & 2.89 & $\pm$ & 0.09 \\
$m_u$ [GeV] & 0.00127 & $\pm$ & 0.00046 \\
$m_c$ [GeV] & 0.619 & $\pm$ & 0.084 \\
$m_t$ [GeV] & 171.7 & $\pm$ & 3.0 \\
$\sin \theta^q_{12}$ & 0.2246 & $\pm$ & 0.0011 \\
$\sin \theta^q_{23}$ & 0.042 & $\pm$ & 0.0013 \\
$\sin \theta^q_{13}$ & 0.0035 & $\pm$ & 0.0003 \\
$\dckm$ & 1.2153 & $\pm$ & 0.0576 \\
$\lambda$ & 0.521 & $\pm$ & 0.01 \\
\bottomrule
\end{tabular}
} 
\begin{tabular}{@{}cr@{ }c@{ }l@{}}
\toprule
Observable  & \multicolumn{3}{c}{Exp. value} \\
\midrule
$\Delta m_\odot^2$ [eV$^2$] & ($7.5$ & $\pm$ & $0.185$)$\times 10^{-5}$ \\
$\Delta m_{\rm 31}^2$ (NH) [eV$^2$] & ($2.47$ & $\pm$ & $0.0685$)$\times 10^{-3}$ \\
$\Delta m_{\rm 31}^2$ (IH) [eV$^2$] & ($-2.355$ & $\pm$ & $0.0540$)$\times 10^{-3}$ \\
$\sin^2 \theta^l_{12}$ & 0.30 & $\pm$ & 0.013 \\
$\sin^2 \theta^{l,\text{(NH \& IH1)}}_{23}$   & 0.41 & $\pm$ & 0.031 \\
$\sin^2 \theta^{l, \text{(IH2)}}_{23}$ & 0.59 & $\pm$ & 0.022 \\
$\sin^2 \theta^l_{13}$ & 0.023 & $\pm$ & 0.0023 \\
$m_e$ [MeV] & 0.48657 & $\pm$ & $0.02433$ \\
$m_{\mu}$ [GeV] & 0.10272 & $\pm$ & 0.00514 \\
$m_{\tau}$ [GeV] & 1.74624 & $\pm$ & 0.08731 \\
\bottomrule
\end{tabular}
\caption{Experimental values of observables at $\mu = M_Z$ used for our fits. 
The quark and charged lepton masses are taken from Ref.\ \cite{Xing:2007fb}, 
quark mixing parameters from Ref.\ \cite{Joshipura:2011nn}, neutrino mixing 
parameters from Ref.\ \cite{GonzalezGarcia:2012sz} (table 1, second column). 
A $5\,\%$ uncertainty is assumed for the charged leptons, as explained in the text. The value of the Higgs quartic coupling
$\lambda$ is derived from the measurements of ATLAS \cite{atlas:2012gk} and 
CMS \cite{cms:2012gu} as explained in \secref{higgsmass}. Note that in our convention the Higgs self-interaction term in the Lagrangian is 
$- \frac{\lambda}{4} (\phi^\dagger \phi)^2$.}
\label{tab:expvals}
\end{table}
\begin{table}[t]
\centering
\begin{tabular}{@{}cc@{}}
\toprule
Obs.  & Value [GeV] \\
\midrule
$m_d$ & 0.00114 $\pm$ 0.000495 \\
$m_s$ & 0.022 $\pm$ 0.0065 \\
$m_b$ & 1.0 $\pm$ 0.04 \\
\bottomrule
\end{tabular} \,
\begin{tabular}{@{}cc@{}}
\toprule
Obs.  & Value [GeV]  \\
\midrule
$m_u$ & 0.00048 $\pm$ 0.000185 \\
$m_c$ & 0.235 $\pm$ 0.0345 \\
$m_t$ & 74.0 $\pm$ 3.85 \\
\bottomrule
\end{tabular} \,
\begin{tabular}{@{}cc@{}}
\toprule
Obs.  & Value [GeV]  \\
\midrule
$m_e \times 10^3$ & 0.46965 $\pm$ 0.02348 \\
$m_{\mu}$ & 0.09915 $\pm$ 0.00496 \\
$m_{\tau}$ & 1.68558 $\pm$ 0.08428 \\
\bottomrule
\end{tabular}
\caption{Experimental values of observables at $\mu = \mgut$ \cite{Joshipura:2011nn} 
used for non-SUSY fits without RGE. For mixing parameters as well as neutrino 
mass-squared differences the same values as in \tabref{expvals} are used. 
A $5\,\%$ uncertainty is assumed for the charged leptons, as explained in the text.}
\label{tab:expvalsnorge}
\end{table}

For the minimization we use the downhill simplex algorithm \cite{Nelder:1965,Press:2007} in its implementation from the GNU Scientific Library \cite{Glassi:2009}, which also provides useful functions for numerical matrix diagonalization and for solving differential equations numerically. 
The parallelized computations are performed on the computer cluster of the Max-Planck-Institut f\"ur Kernphysik, Heidelberg, where up to 1700 CPU cores can be used.

Let us stress a general caveat of numerical minimization. The problem at hand is non-linear and 
multidimensional -- therefore many local minima exist. With numerical algorithms it is impossible to determine
whether a minimum is a global minimum of the function under consideration. 
A standard procedure to increase the confidence that a global minimum
out of the many local ones has been found is to start the minimization many times with different initial parameters and to
choose the lowest out of the many local minima that will be found. Furthermore one 
can perturb a minimum and restart the minimization from the perturbed 
point \cite{Press:2007,Glassi:2009}. 
These steps can be repeated many times until no improvement of the minimum is found any more. 
Both methods are used by our program. 

\begin{table}[t]
\centering
\begin{tabular}{@{}cr@{ }c@{ }lr@{ }c@{ }lr@{ }c@{ }l@{}}
\toprule
Observable  & \multicolumn{3}{c}{$\tan \beta = 50$} & \multicolumn{3}{c}{$\tan \beta = 38$} & \multicolumn{3}{c}{$\tan \beta = 10$} \\
\midrule
$m_u/m_c$ & 0.0027 & $\pm$ & 0.0006 & 0.0027 & $\pm$ & 0.0006 & 0.0027 & $\pm$ & 0.0006 \\
$m_d/m_s$ & 0.051 & $\pm$ & 0.007 & 0.051 & $\pm$ & 0.007 & 0.051 & $\pm$ & 0.007 \\
$m_c/m_t$ & 0.0023 & $\pm$ & 0.0002 & 0.0024 & $\pm$ & 0.0002 & 0.0025 & $\pm$ & 0.0002 \\
$m_s/m_b$ & 0.016 & $\pm$ & 0.002 & 0.017 & $\pm$ & 0.002 & 0.019 & $\pm$ & 0.002 \\
$m_e/m_\mu$ & 0.0048 & $\pm$ & 0.0002 & 0.0048 & $\pm$ & 0.0002 & 0.0048 & $\pm$ & 0.0002 \\
$m_\mu/m_\tau$ & 0.05 & $\pm$ & 0.002 & 0.054 & $\pm$ & 0.002 & 0.059 & $\pm$ & 0.002 \\
$m_b/m_\tau$ & 0.73 & $\pm$ & 0.04 & 0.73 & $\pm$ & 0.04 & 0.73 & $\pm$ & 0.03 \\
$\sin \theta^q_{12}$ & 0.227 & $\pm$ & 0.001 & 0.227 & $\pm$ & 0.001 & 0.227 & $\pm$ & 0.001 \\
$\sin \theta^q_{23}$ & 0.0371 & $\pm$ & 0.0013 & 0.0386 & $\pm$ & 0.0014 & 0.04 & $\pm$ & 0.0014 \\
$\sin \theta^q_{13}$ & 0.0033 & $\pm$ & 0.0007 & 0.0035 & $\pm$ & 0.0007 & 0.0036 & $\pm$ & 0.0007 \\
$\dckm$ & 0.9828 & $\pm$ & 0.1784 & 0.9828 & $\pm$ & 0.1784 & 0.9828 & $\pm$ & 0.1787 \\
$\Delta m_{21}^2 / \Delta m_{31}^2$ & 0.03036 & $\pm$ & 0.0011 & 0.03036 & $\pm$ & 0.0011 & 0.03036 & $\pm$ & 0.0011 \\
$\Delta m_{31}^2$ (NH) [eV$^2$] & (2.47  & $\pm$ & 0.0685)$\times 10^{-3}$ & (2.47  & $\pm$ & 0.0685)$\times 10^{-3}$ & (2.47  & $\pm$ & 0.0685)$\times 10^{-3}$\\
$m_\tau$ [GeV] & 0.773 & $\pm$ & 0.0387 & 0.950 & $\pm$ & 0.0475 & 1.022 & $\pm$ & 0.0511 \\
$m_t$ [GeV] & 94.7 & $\pm$ & 9.4 & 94.7 & $\pm$ & 9.4 & 92.2 & $\pm$ & 8.7 \\
\bottomrule
\end{tabular}
\caption{Experimental values of observables at $\mu = \mgut$ \cite{Joshipura:2011nn} 
used for SUSY fits without RGE. The ratio of solar to atmospheric neutrino mass-squared 
difference is calculated from their values at $\mu = \mz$ as given in \tabref{expvals}. 
The top-quark mass $m_t$ and the tau mass $m_\tau$ at $\mu = \mgut$ for $\tan \beta = 50, \, 10$ are taken from Ref.~\cite{Xing:2007fb}, the values for $\tan \beta = 38$ are interpolations.
For the neutrino mixing angles as well as $\Delta m_{\rm 31}^2$ the values at $\mu = \mz$ as given in \tabref{expvals} are used.}
\label{tab:expvalsnorgesusy}
\end{table}
\subsection{Neutrino Data}
\label{sec:nhvsih}
In the neutrino sector the absolute mass scale is experimentally not yet determined. At present, only the solar mass-squared difference $\Delta m_{21}^2$ and the absolute value of the atmospheric mass-squared difference $|\Delta m_{31}^2|$ are known \cite{GonzalezGarcia:2012sz},
\begin{align}
\Delta m_{21}^2 & = 7.5 \pm 0.185 \times 10^{-5} \text{ eV}^2 \notag \, , \\ 
\Delta m_{31}^2 & = 2.47 \pm 0.0685 \times 10^{-3} \text{ eV}^2 \; \text{(NH)} \, , \notag \\
\Delta m_{31}^2 & = -2.355 \pm 0.0540 \times 10^{-3} \text{ eV}^2 \; \text{(IH)} \, , \notag
\end{align}
where NH (normal hierarchy) and IH (inverted hierarchy) refer to two currently viable situations with $m_1 < m_2 < m_3$ and $m_3 < m_1 < m_2$, respectively. 
Furthermore, we symmetrized the uncertainties, as explained before.

Besides the different signs and values of $\Delta m_{31}^2$, also the neutrino mixing parameters have different preferred values depending on which mass hierarchy is assumed \cite{Tortola:2012te,Fogli:2012ua,GonzalezGarcia:2012sz}. This hierarchy dependence is mostly pronounced for $\sin^2 \theta_{23}^l$. Here, the best-fit value of $\sin^2 \theta_{23}^l$ depends on aspects of the analysis, including the experiments that were considered. Comparing Refs.\ \cite{Tortola:2012te,Fogli:2012ua,GonzalezGarcia:2012sz} we notice that there currently exist two different equally valid best-fit values and corresponding $1\,\sigma$ regions for $\sin^2 \theta_{23}^l$. We take the values to which the models will be fitted from Ref.\ \cite{GonzalezGarcia:2012sz} and distinguish the following cases in our analysis:
\begin{align}
\sin^2 \theta_{23}^l & = 0.41 \pm 0.031	\quad \text{NH}		\, ,	\notag	\\
\sin^2 \theta_{23}^l & = 0.41 \pm 0.031	\quad \text{IH1}	\, ,			\\
\sin^2 \theta_{23}^l & = 0.59 \pm 0.022	\quad \text{IH2}	\, .	\notag
\end{align}
The quality of fits with the inverted neutrino mass hierarchy had the same quality for both IH1 and IH2. Hence, we will stick in our discussion of results in \secref{results} to the case IH2. While the other 
groups performing global neutrino fits do not have these two solutions, we decided to 
use the results from Ref.\ \cite{GonzalezGarcia:2012sz}, in order to have a test for 
the octant-dependence of the $SO(10)$ fit results.

\newpage
\subsection{Higgs Mass and Quartic Coupling}
\label{sec:higgsmass}
Although the Higgs boson mass \mhiggs does not enter the \soten relations in \eqnref{mrel} there is
interplay between \mhiggs and the fermion observables during renormalization group
evolution. In RGE in the non-supersymmetric case the Higgs quartic coupling $\lambda$ appears, which in the SM is 
related to \mhiggs by\footnote{In our convention the Higgs self-interaction term in the Lagrangian is 
$- \frac{\lambda}{4} (\phi^\dagger \phi)^2$.}
\beq
\label{eqn:lambdavshiggs}
\lambda = \frac{2}{v^2} \, m_\text{H}^2 \; .
\ee
Recently, the ATLAS and CMS experiments at the Large Hadron Collider (LHC) have observed a new particle,  
which is in good agreement with a Standard Model Higgs boson, with the mass \cite{atlas:2012gk,cms:2012gu}
\begin{equation}
\label{eqn:lhchiggsmass}
\begin{aligned}
m_\text{H} & = 126.0 \pm 0.4 \text{ (stat)} \pm 0.4 \text{ (sys)} \text{ GeV \hspace{0.5cm} (ATLAS)} \\
m_\text{H} & = 125.3 \pm 0.4 \text{ (stat)} \pm 0.5 \text{ (sys)} \text{ GeV \hspace{0.5cm} (CMS)} \; .
\end{aligned}
\end{equation}
For our analysis we take a conservative estimate of the true Higgs mass, since there is no official
combined analysis available. Our 1\,$\sigma$ interval shall
overlap exactly the 1\,$\sigma$ intervals of the ATLAS and CMS experiments and we take the central
value of this range as best-fit point. Thus, we take for our fits
\beq
\label{eqn:myhiggsmass}
m_\text{H} = 125.6 \pm 1.2 \text{ GeV} \, .
\ee
The standard error propagation
formula applied to \eqnref{lambdavshiggs} then yields
\beq
\lambda = 0.521 \pm 0.010 \; .
\ee
Note that for fits at \mgut, i.e.~without RG evolution, we do not take into account the Higgs mass, since in that case there is no restriction on $\lambda$ from the other observables.
\subsubsection*{Supersymmetric Case}
Above the supersymmetry breaking scale \msusy, supersymmetry fixes $\lambda$ to 
be\footnote{We apply GUT normalization to the $\U(1)_\mathrm{Y}$ charge.} \cite{Haber:1996fp}
\beq
\lambda(\mu \geq \msusy) = \frac 14 \left( \frac 35 \, g_1^2 + g_2^2 \right) ( \mu ) \, .
\ee
Below \msusy the Higgs mass receives radiative corrections, the leading one given in a rough
approximation (within the MSSM) by \cite{Haber:1996fp}
\beq
\label{eqn:higgsrad}
\mhiggs^2 = \mz^2 + \frac{ 3 \, g_2^2 \, m_t^4(\mu_t) }{ 8 \pi^2 M_W^2
} \, \text{ln} \left( \frac{\msusy^2}{m_t^2(\mu_t) } \right)  ,
\ee
with $\mu_t = \sqrt{m_t \, \msusy}$ and all SUSY particles are assumed to have masses 
around \msusy in this approximation. By varying \msusy one can reproduce the measured 
value of \mhiggs as given in \eqnref{lhchiggsmass,myhiggsmass}. Solving \eqnref{higgsrad} for
\msusy yields $\msusy \approx$~1~TeV.
Since our main goal is to fit fermion masses and mixings within the \soten framework
and not performing a detailed analysis of the MSSM, we do not specify 
\msusy or the specific SUSY spectrum.
Hence, we will not try to fit \mhiggs in the supersymmetric models.
%
%
%
\subsection{\label{sec:rge}Renormalization Group Evolution}
The relations in \eqnref{yrel} have to be obeyed at \mgut. Therefore, for a given set of \soten parameters, in order to calculate the model predictions for the observables one has to use RGEs and evolve the parameters down to the energy scale at which the observables are known. In addition one has to integrate out heavy degrees of freedom during this process at their mass scale. In our case this applies to the right-handed neutrinos ($\nu_{R_i}$), as their masses usually lie somewhere between $10^{10}$~GeV and \mgut. After integrating out a degree of freedom one ends up with an effective field theory (EFT) and has to match coefficients of effective operators with parameters from the full theory. Since $\nu_{R_i}$ are not degenerate in general, one has to integrate out several times and thus has to use different EFTs during the evolution from \mgut to \mz. This formalism is nicely described in Refs.~\cite{Antusch:2002rr, Antusch:2005gp}. We use the 1-loop $\beta$-functions as presented in Ref.~\cite{Antusch:2005gp} for the SM and MSSM, respectively (see also \cite{Chankowski:2001mx}). The beta-functions are also presented in appendix~\ref{sec:betafcts} for reference. We should mention that we do not integrate out the top-quark below $\mu = m_t(m_t)$, since the energy scales $m_t$ and \mz are quite close. Furthermore we assume $M_{\text{SUSY}} = M_{Z}$, i.e.~we use the beta-functions of the MSSM for the evolution of parameters down to \mz in case of SUSY models, since this is just a small effect as long as the SUSY breaking scale is not too far away from \mz. We expect models being able to fit experimental data with $M_{\text{SUSY}} = M_Z$ to be equally well suited to fit the data with $M_{\text{SUSY}} = 500$~GeV or 1~TeV. Finally, since we do not specify the SUSY spectrum, we also do not consider SUSY threshold effects, which may have an impact for large $\tan \beta$ \cite{Hall:1993gn,Carena:1994bv,Hempfling:1993kv,Blazek:1995nv,Freitas:2007dp,Antusch:2008tf,
Ross:2007az,Joshipura:2011nn}.

\section{\label{sec:results}Results}
In this section we present and discuss the results of our analysis. We quantify the deviation of model predictions by stating the pulls of all the observables considered. The pull of a model with respect to an observable $y_i$ is defined as
\be 
\mbox{pull}(y_i) = \frac{y_i^{\rm theo}(x) - y_i^{\rm exp}}{\sigma_i^{\rm exp}}  ,
\ee
with the variables as defined in \eqnref{chisqdef} on page \pageref{eqn:chisqdef}. The pull measures the deviation of theoretical predictions (or best-fit values) from experimentally measured values in units of uncertainty of the observable. Its sign shows whether the theoretical prediction is too small or too large.

\subsubsection*{Different Sets of Observables}
We will present the results of fits including RGE, where the input values are taken at $\mu = M_Z$, as well as the results of fits made without RGE, as explained in Sec.\ \ref{sec:introsoten}. Our full set of observables to which the models are fitted are the masses of quarks and charged leptons, mass-squared differences of neutrinos, mixing angles of quarks and leptons, the $CP$ phase $\dckm$ in the CKM matrix and the Higgs quartic coupling $\lambda$. The full set of observables is used only in non-SUSY models with full RG analysis, $\lambda$ is generally not fitted in SUSY models and in non-SUSY models without RG analysis (see \secref{higgsmass} for details).   Hence the number of observables taken into account for the fits is 18 or 19. Numerical input values can be found in \secref{fittingprocedure}. There we also pointed out that in case of the inverted neutrino mass hierarchy currently two numerically different best-fit solutions exist for the value of $\sin^2 \theta_{23}^l$. Since good fits could be achieved for both possibilities, we restrict the discussion of our results to the case where $\sin^2 \theta_{23} ^l > 0.5$. For the normal hierarchy, the experimental best-fit value is $\sin^2 \theta_{23} ^l = 0.41 < 0.5$.
We now proceed with the discussion of each model.

\subsection{Minimal Model with $10_H + \overline{126}_H$ (MN, MS)}
\label{sec:resmin}

\begin{table}[t]
\centering
\begin{tabular}{@{}rclcc@{}} 
\toprule
Model	&	$\tan \beta$	&  Comment	& $\chi^2_{\rm NH}$	& $\chi^2_{\rm IH}$ \\
\midrule
MN		&		--		&  no RGE	&	1.103	&	395	\\ 
			&		--	&  RGE		&	22.97	&	680	\\ 
\addlinespace
MS		&		50		&  no RGE	&	9.411	&	200	\\ 
			&		50	&  RGE		&	3.294	&	420	\\ 
\addlinespace
\addlinespace
			&		38	&  no RGE	&	9.715	&	300	\\ 
			&		38	&  RGE		&	3.016	&	--	\\ 
\addlinespace
			&		10	&  no RGE	&	10.45	&	225	\\ 
			&		10	&  RGE		&	2.889	&	--	\\
\bottomrule
\end{tabular}
\caption{Fit results for models MN and MS (Higgs content $10_H + \overline{126}_H$, 19 free parameters), to 
19 and 18 observables, respectively.  
We differentiate between fits to observables without RGE (``no RGE'') 
and fits to observables including RGE.  For the fits 
of MN including RGE the Higgs quartic coupling has also been fitted, 
as described in \secref{higgsmass}. No fits are made for the
time-consuming cases with RGE in the inverted hierarchy for $\tan
\beta = 38$ and $10$, because for NH the fits give essentially
identical $\chi_{\rm min}^2$-values.}
\label{tab:chisqresRgeYb}
\end{table}

\begin{table}[t]
\centering
\begin{tabular}{@{}ccccc@{}}
\toprule
			&	\multicolumn{2}{c}{MN, no RGE} &	\multicolumn{2}{c}{MN, with RGE} \\
\cmidrule(l){2-3} \cmidrule(l){4-5}
Observable  &  best-fit  &  pull  &  best-fit  &  pull   \\
\midrule
$m_d$ & 0.00067 & -0.9458 & 0.00298 & 0.0621   	\\
$m_s$ & 0.02406 & 0.3172 & 0.06887 & 0.8951 	\\
$m_b$ & 1.00309 & 0.0772 & 2.89370 & 0.0411		\\
$m_u$ & 0.00048 & 0.0072 & 0.00131 & 0.0977 	\\
$m_c$ & 0.24243 & 0.2153 & 0.70754 & 1.0541 	\\
$m_t$ & 73.6931 & -0.0797 & 161.411 & -3.4295 	\\
$\sin \theta^q_{12}$ & 0.22462 & 0.0227 & 0.22476 & 0.1433 	\\
$\sin \theta^q_{23}$ & 0.04204 & 0.0304 & 0.04170 & -0.2291	\\
$\sin \theta^q_{13}$ & 0.00350 & 0.0091 & 0.00342 & -0.2520	\\
$\dckm$ & 1.21930 & 0.0699 & 1.25285 & 0.6525	\\
$\Delta m_{\rm 21}^2$ & 7.50$\times 10^{-5}$ & 0.0180 & 7.53$\times 10^{-5}$ & 0.1626	\\
$\Delta m_{31}^2$ & 2.47$\times 10^{-3}$ & -0.0204 & 2.46$\times 10^{-3}$ & -0.1858	\\
$\sin^2 \theta^l_{12}$ & 0.30039 & 0.0303 & 0.29864 & -0.1044	\\
$\sin^2 \theta^l_{23}$ & 0.40631 & -0.1189 & 0.34571 & -2.0739	\\
$\sin^2 \theta^l_{13}$ & 0.02262 & -0.1652 & 0.01847 & -1.9678	\\
$m_e$ &  4.697$\times 10^{-4}$  &  ---  & 0.00049 & 0.0704	\\
$m_{\mu}$ &  9.914$\times 10^{-2}$  &  ---  & 0.10143 & -0.2508	\\
$m_{\tau}$ &  1.686  &  ---  & 1.73804 & -0.0939	\\
$\lambda$ &  ---  &  ---  & 0.52731 & 0.6307	\\
\midrule
$\chi^2_{\rm min}$ & 	 & {\bf 1.103 } & 	 & {\bf 22.97 }	\\
\bottomrule
\end{tabular}
\caption{19 Observables and pulls for model MN (minimal non-SUSY, $10_H+\overline{126}_H$, 
19 free parameters) with and without considering RGE.  Masses are given in GeV, mass-squared differences in eV$^2$.}
\label{tab:m1ntresults}
\end{table}

%

Let us start by comparing our program to previous fit results. 
In case of no RGE, our minimal $\chi^2$-value in the non-SUSY case is 1.1, to be compared with the value $\chi^2_ {\rm min}  \approx 0.7$ in Ref.\  \cite{Joshipura:2011nn}. 
The SUSY model MS with normal neutrino mass hierarchy has been
analyzed before (without RG evolution)
\cite{Bertolini:2006pe,Joshipura:2011nn}, albeit with older data
underlying the analyses. The results lie in the range between $\chi^2_{\rm min}
= 3.7$ and $\chi^2_{\rm min}= 5.1$ in case of type I seesaw dominance. Our fits
yield $\chi^2_{\rm min} =$ 9.41, 9.72, and 10.45 for $\tan \beta =$ 50, 38, and 10, respectively. The results are summarized in \tabref{chisqresRgeYb} and the best-fit values of observables and corresponding pulls, to be discussed below, are compiled in \tabref{m1ntresults,msntRgeNoYb}.

With inverted neutrino mass hierarchy it was impossible to produce a
good fit of this model, with $\chi^2_{\rm min} >$ 200 (400) in case of
SUSY (non-SUSY) models. Within SUSY versions of such models this
observation has already been made by other authors
\cite{Bertolini:2006pe}, however without including full RGE during the
fitting procedure and hence with neglecting running of neutrino parameters which can be sizable in the inverted hierarchy case. Therefore, our conclusion is more stringent. Since with the inverted hierarchy it is impossible to fit the data, we present only results for the normal neutrino mass hierarchy.

\begin{figure}[tb]
\begin{center}
\epsfig{file=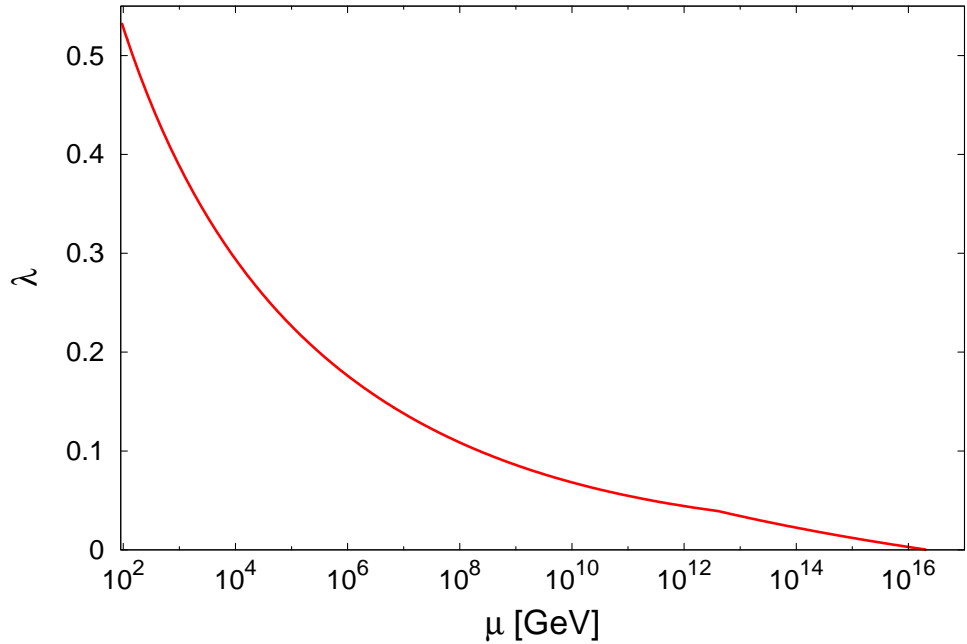,width=0.8\linewidth}
\caption{\label{fig:lamrunm1nt}Running of $\lambda$ for the best-fit parameters of model MN with normal neutrino mass hierarchy and RG evolution. 
The kink between $\mu = 10^{12} \text{ GeV}$ and $\mu = 10^{13} \text{
  GeV}$ results from integrating out the heaviest right-handed neutrino. }
\end{center}
\end{figure}
\begin{table}[t]
\centering
\begin{tabular}{@{}ccccccc@{}}
\multicolumn{7}{l}{MS (with RGE)} \\
\toprule
			&	\multicolumn{2}{c}{$\tan \beta = 50$} &  \multicolumn{2}{c}{$\tan \beta = 38$}  &  \multicolumn{2}{c}{$\tan \beta = 10$} \\ \cmidrule(l){2-3} \cmidrule(l){4-5} \cmidrule(l){6-7}
Observable  &  best-fit  &  pull  &  best-fit  &  pull  &  best-fit  &  pull \\
\midrule
$m_d$ & 0.00087 & -1.6714 & 0.00090 & -1.6449 & 0.00091 & -1.6381  \\
$m_s$ & 0.04512 & -0.6371 & 0.04711 & -0.5089 & 0.04870 & -0.4063  \\
$m_b$ & 2.87626 & -0.1526 & 2.88217 & -0.0870 & 2.88499 & -0.0557  \\
$m_u$ & 0.00127 & 0.0018 & 0.00127 & 0.0068 & 0.00127 & 0.0064  \\
$m_c$ & 0.62848 & 0.1129 & 0.62738 & 0.0997 & 0.62854 & 0.1135  \\
$m_t$ & 171.453 & -0.0823 & 171.522 & -0.0593 & 171.539 & -0.0537  \\
$\sin \theta^q_{12}$ & 0.22460 & -0.0040 & 0.22460 & -0.0018 & 0.22460 & -0.0009  \\
$\sin \theta^q_{23}$ & 0.04191 & -0.0675 & 0.04193 & -0.0565 & 0.04193 & -0.0543  \\
$\sin \theta^q_{13}$ & 0.00351 & 0.0241 & 0.00351 & 0.0322 & 0.00351 & 0.0314  \\
$\dckm$ & 1.21318 & -0.0364 & 1.21398 & -0.0225 & 1.21409 & -0.0205  \\
$\Delta m_{\rm 21}^2$ & 7.50$\times 10^{-5}$ & 0.0021 & 7.50$\times 10^{-5}$ & 0.0013 & 7.50$\times 10^{-5}$ & 0.0009  \\
$\Delta m_{\rm 31}^2$ & 2.47$\times 10^{-3}$ & -0.0022 & 2.47$\times 10^{-3}$ & -0.0013 & 2.47$\times 10^{-3}$ & -0.0010  \\
$\sin^2 \theta^l_{12}$ & 0.30015 & 0.0112 & 0.30007 & 0.0053 & 0.30004 & 0.0028  \\
$\sin^2 \theta^l_{23}$ & 0.40960 & -0.0129 & 0.40977 & -0.0073 & 0.40987 & -0.0043  \\
$\sin^2 \theta^l_{13}$ & 0.02299 & -0.0045 & 0.02297 & -0.0138 & 0.02297 & -0.0123  \\
$m_e$ & 0.00049 & 0.0702 & 0.00049 & 0.0660 & 0.00049 & 0.0605  \\
$m_{\mu}$ & 0.10315 & 0.0839 & 0.10306 & 0.0665 & 0.10298 & 0.0513  \\
$m_{\tau}$ & 1.76204 & 0.1809 & 1.75740 & 0.1278 & 1.75528 & 0.1035  \\
\midrule
$\chi^2_{\rm min}$ & 	 & {\bf 3.294 } & 	 & {\bf 3.016 } & 	 & {\bf 2.889 } \\
\bottomrule
\end{tabular}
\caption{18 Observables and pulls for model MS (minimal SUSY, $10_H+\overline{126}_H$, 19 free parameters) with RGE, for different
values of $\tan \beta$. Masses are given in GeV, mass-squared differences in eV$^2$.}
\label{tab:msntRgeNoYb}
\end{table}

We see that for the minimal models in the non-supersymmetric case, including the full RG analysis worsens the fit considerably, while doing the same for the SUSY model gives a better result than fitting without RG evolution (\tabref{chisqresRgeYb}). In case of non-SUSY models there is an additional constraint when fitting with RG evolution, since in that case we also consider the Higgs mass (see \secref{higgsmass}). 
Still, both non-SUSY and SUSY models can fit the data, the SUSY models being in better agreement. For the SUSY model we see no preferred value of $\tan \beta$ from our fits. 

Let us now discuss the different contributions to $\chi^2_{\rm min}$.
We show the best-fit values of observables and their corresponding pulls in \tabref{m1ntresults,msntRgeNoYb}. 
In case of non-SUSY fits without RGE one observes that the dominating contribution to $\chi_{\rm min}^2$ is due to the pull of the down-quark mass. 
In case of SUSY fits without RGE, we fit for the sake of better comparison with 
Ref.\ \cite{Joshipura:2011nn} mass ratios instead of masses.  
There the dominating contribution is the $m_d/m_s$ ratio. 
In case of non-SUSY fits the main contribution to $\chi^2_{\rm min}$ comes from
the mass of the top-quark ($\sim 3.4  \,\sigma$),  
followed by the pulls of $\sin^2 \theta_{23}^l$ and $\sin^2 \theta_{13}^l$. \\

{\bf Higgs mass vs.~top-quark mass:}\
The tension in the fit due to the top-quark mass is easily understood from the relatively light Higgs mass (and hence low quartic coupling $\lambda$). Namely, as well-known from the vacuum stability problem  
(for recent analyses, see Refs.\ \cite{Holthausen:2011aa,EliasMiro:2011aa,Xing:2011aa,Degrassi:2012ry,Masina:2012tz}), 
the beta-function governing the evolution of the Higgs quartic coupling $\lambda$, as given in appendix \ref{sec:betafctsSM}, is dominated by the top-quark Yukawa, and drives $\lambda$ towards negative values 
when going from low to high scale. 
Note that in 
our fit we start with $\lambda \ge 0$ at the GUT scale, and hence do not have a problem of negative 
$\lambda$. Nevertheless, the ``too large'' top-quark mass causes some pressure on the fits. 
The large top-quark Yukawa coupling\footnote{We neglect for this argument 
effects of higher loop or other corrections to the Higgs potential.  
Moreover, Dirac Yukawas have a similar, though somewhat weaker, effect than the top-quark Yukawa, 
see e.g.\ \cite{Casas:1999cd,Rodejohann:2012px}.}  
favors a larger Higgs mass than experimentally established. This ultimately results in 
relatively large pulls for the top-quark mass in fits with RG evolution. 
Therefore, we also performed a fit of model MN without including the Higgs mass. 
As expected, the fit improves from $\chi^2_{\rm min} = 22.97$ to $\chi^2_{\rm min} = 8.21$.

\begin{table}[t]
\centering
\begin{tabular}{@{}ccc@{}}
\toprule
MN (NH, RGE)	&	with Higgs		&	w/o Higgs		\\
\midrule
heavy top		&	22.97	&	8.21	\\
light top		&	10.06	&	6.70		\\
\bottomrule
\end{tabular}
\quad
\begin{tabular}{@{}ccc@{}}
\toprule
FN (IH, RGE)	&	with Higgs		&	w/o Higgs		\\
\midrule
heavy top		&	13.3	&	0.67	\\
light top		&	0.98		&	0.50		\\
\bottomrule
\end{tabular}
\caption{Impact of top-quark mass and Higgs mass on fit results for models MN (normal hierarchy) and FN (inverted hierarchy). The \chisq-minima for the fit with a heavy top-quark mass and a light top-quark mass in combination with and without fitting the Higgs mass are shown. }
\label{tab:lighttop}
\end{table}

Recall, however, that there is a discussion on the correct value of the top-quark mass as 
determined via kinematic reconstruction \cite{Langenfeld:2009wd,Hoang:2011TRSlides,Hoang:2008xm}. The top-quark mass determination at the TeVatron is based on the final state of the decay products. Another possibility is to reconstruct the top-quark mass from the total cross section in the top-quark pair production. This method is more rigorous from a theoretical perspective and yields a smaller top-quark mass ($168.9 \pm 3.5$~GeV\footnote{Note that this is the pole-mass $m_t(m_t)$,  while the value given in \tabref{expvals} is the mass at \mz in the $\overline{MS}$ scheme. Converting to $\overline{MS}$ using 1-loop matching~\cite{Hempfling:1994ar} conditions and evolving to $\mu = \mz$ yields $m_t = 158.5 \pm 3.2$~GeV as the observable to be used in the fits with light top-quark mass.} \cite{Langenfeld:2009wd}) than the world average and has larger error bars. 
One can expect that the fit will improve when we use this lower top-quark mass. 
Indeed, $\chi^2_{\rm min}$ reduces from 22.97 to 10.06. 
We summarize different fits with heavy and light top-quark in combination with and 
without Higgs in \tabref{lighttop}. 
Let us remark again that intermediate scales and other details of the full scalar sector are neglected 
in our analysis. Hence, a more correct but model-dependent treatment of the evolution of 
$\lambda$ might lead to perfect compatibility of the measured Higgs mass and large 
top-quark mass (or make the problem worse). 

The evolution of $\lambda$ with energy for the best-fit parameters of model MN with normal neutrino mass hierarchy and RG evolution is shown in Fig.~\ref{fig:lamrunm1nt}. Notice the kink between $\mu = 10^{12} \text{ GeV}$ and $\mu = 10^{13} \text{ GeV}$, which results from integrating out the heaviest neutrino with a mass of $M_3 \simeq 3.6 \times 10^{12} \text{ GeV}$ (see \secref{predictions}). 
There is no further such kink at energies where the other heavy neutrinos are integrated out, since their contribution to the running of $\lambda$ is negligible compared to the contribution of the top-quark.

\subsection{Alternative Minimal Model with $\overline{126}_H + 120_H$}
This model was analyzed only in the non-SUSY version (as it was originally proposed to be attractive in that case \cite{Bajc:2005zf}) and found to be unable to reproduce fermion masses and mixings. Although this result has been obtained previously \cite{Joshipura:2011nn}, 
only the normal hierarchy was considered, and no detailed RGE analysis was performed. 
In this work we used full RGE to arrive at our results, thus our conclusion is stronger. 
Further, we also checked and excluded the possibility of the inverted hierarchy.
The supersymmetric version, that to the best of our knowledge has not been analyzed before, 
is also not an option to save this model (to safe CPU-time, we only fitted the case of $\tan \beta = 10$).  
We find large \chisq-values for this case as well and therefore we only present a table 
with the large \chisq-values, see \tabref{M2}. 
We note however that the normal neutrino mass hierarchy has significantly smaller, though still 
too large \chisq-values.  

\begin{table}[tb]
\centering
\begin{tabular}{@{}rlcc@{}} 
\toprule
	$\tan \beta$	&  Comment				&	$\chi^2_{\rm NH}$	& $\chi^2_{\rm IH}$ \\
\midrule
		--			&  no RGE	&	$103$	&	$910$	\\ 
		--			&  RGE	&	200		&	3859	\\ 
\addlinespace
		10			&  no RGE	&	$247$		&	$1861$	\\ 
			10			&  RGE	& 224	&  4358	  \\ 
\bottomrule
\end{tabular}
\caption{Fit results of an alternative minimal model with $\overline{126}_H + 120_H$ 
Higgs representations, having 17 free parameters. 
No acceptable fit 
was found. }
\label{tab:M2}
\end{table}

\subsection{Model with $10_H + \overline{126}_H + 120_H$ (FN, FS)}

\begin{table}[tb]
\centering
\begin{tabular}{@{}rclcc@{}} 
\toprule
Model		&	$\tan \beta$	&  Comment				&	$\chi^2_{\rm NH}$	&	$\chi^2_{\rm IH}$ \\
\midrule
FN		&		--			&  no RGE	&	$6.6 \times 10^{-5}$	&	$2.5 \times 10^{-4}$	\\ 
			&		--			&  RGE	&	11.2		&	13.3	\\ 
\addlinespace
FS		&		50			&  no RGE	&	$9.0 \times 10^{-10}$		&	$3.9 \times 10^{-8}$	\\ 
			&		50			&  RGE	&	$6.9 \times 10^{-6}$	&  0.602	  \\ 
\bottomrule
\end{tabular}
\caption{Values of $\chi^2_{\rm min}$ at the best-fit position for the model with $10_H + \overline{126}_H + 120_H$ (18 free parameters) in case of normal (NH) and inverted (IH) neutrino mass hierarchy. For IH we present the solution with $\sin^2 \theta_{23}^l > 0.5$, but both possibilities yield equally good fits. Remarks as in \tabref{chisqresRgeYb} apply analogously. Model FN (FS) contains 19 (18) free parameters.}
\label{tab:chisqfullmodel}
\end{table}

This class of models, in spite of having one more Higgs representation, through the additional constraints (i.e.~assuming spontaneous $CP$ violation, see \secref{models}) has one parameter less than the minimal models. Nevertheless, it is not only able to fit the data, but reproduces the data even much better than the other models. This is especially the case for the SUSY versions of this model. Furthermore, these models are also able to fit the data with inverted neutrino mass hierarchy, which differentiates them clearly from the previous models. Since we do not observe a significant difference in the quality of fits of SUSY models with different values of $\tan \beta$,  we fitted this model only for $\tan \beta = 50$ and $\tan \beta = 10$, which again yield very similar results, as in the case of the $10_H + \overline{126}_H$ model. Therefore we present here only the detailed results for $\tan \beta = 50$. Our results are tabulated in \tabref{chisqfullmodel} and the best-fit values of observables and their pulls are compiled in \tabref{f1ntpulls,fsntpulls}.

Let us first discuss the fits with normal hierarchy.
This setup has been analyzed without RG evolution in the SUSY case in Refs.\ \cite{Grimus:2006rk,Altarelli:2010at,Joshipura:2011nn} with $\chi^2_{\rm min}$ ranging between 0.01 and 0.33. The non-SUSY case has been fitted to data only in Ref. \cite{Joshipura:2011nn} and results in $\chi^2_{\rm min} \sim 10^{-6}$.
Again we observe that fitting the non-SUSY version of this model including RG evolution significantly worsens the fit.  The SUSY fits turn out to be even better than the non-SUSY fits. Here, fits with RGE as well as fits without RGE yield \chisq-values that are essentially zero.

\begin{table}[t]
\centering
\begin{tabular}{@{}ccccc@{}}
\toprule
			&	\multicolumn{2}{c}{FN, NH, RGE} &	\multicolumn{2}{c}{FN, IH, RGE} \\ \cmidrule(l){2-3} \cmidrule(l){4-5}
Observable  &  best-fit  &  pull  &  best-fit  &  pull	\\
\midrule
$m_d$ & 0.00295 & 0.0414 & 0.00304 & 0.1167  \\
$m_s$ & 0.06199 & 0.4512 & 0.06650 & 0.7417  \\
$m_b$ & 2.88874 & -0.0140 & 2.88956 & -0.0049  \\
$m_u$ & 0.00127 & 0.0003 & 0.00127 & 0.0008  \\
$m_c$ & 0.62395 & 0.0590 & 0.62377 & 0.0568  \\
$m_t$ & 161.943 & -3.2525 & 161.207 & -3.4977  \\
$\Delta m_{21}^2$ & 7.50$\times 10^{-5}$ & 0.0015 & 7.50$\times 10^{-5}$ & -0.0001  \\
$\Delta m_{31}^2$ & 2.47$\times 10^{-3}$ & -0.0037 & -2.35$\times 10^{-3}$ & 0.0019  \\
$\sin \theta^q_{12}$ & 0.22460 & -0.0044 & 0.22460 & 0.0042  \\
$\sin \theta^q_{23}$ & 0.04192 & -0.0646 & 0.04182 & -0.1347  \\
$\sin \theta^q_{13}$ & 0.00350 & -0.0031 & 0.00350 & -0.0007  \\
$\delta_{\rm CKM}$ & 1.21402 & -0.0217 & 1.21650 & 0.0213  \\
$\sin^2 \theta^l_{12}$ & 0.30006 & 0.0048 & 0.30000 & 0.0002  \\
$\sin^2 \theta^l_{23}$ & 0.41029 & 0.0093 & 0.59025 & 0.0116  \\
$\sin^2 \theta^l_{13}$ & 0.02302 & 0.0078 & 0.02300 & 0.0015  \\
$m_e$ & 0.00049 & 0.0001 & 0.00049 & 0.0005  \\
$m_{\mu}$ & 0.10232 & -0.0777 & 0.10212 & -0.1173  \\
$m_{\tau}$ & 1.74663 & 0.0045 & 1.74178 & -0.0511  \\
$\lambda$ & 0.52745 & 0.6455 & 0.52792 & 0.6917  \\
\midrule
$\chi^2_{\rm min}$ &       & {\bf 11.2 } &         & {\bf 13.3 } \\
\bottomrule
\end{tabular}
\caption{19 Observables and pulls for model FN (18 free parameters) fitted assuming normal (NH) or inverted (IH) hierarchy. Masses are given in GeV, mass-squared differences in eV$^2$.}
\label{tab:f1ntpulls}
\end{table}

We now turn to the inverted hierarchy. In contrast to the $10_H + \overline{126}_H$ model an inverted neutrino mass hierarchy is viable here. The SUSY case with inverted hierarchy has been fitted to data in Ref.\ \cite{Grimus:2006rk}, giving $\chi^2_{\rm min} = 0.011$, but RGE was not taken into account, 
which is especially important for the inverted hierarchy. 
We are not aware of any analysis of the non-SUSY case with inverted hierarchy. 
In our analysis, in the non-SUSY model the fit quality is approximately the same as in the normal hierarchy. Again, when fitting with RGE, inclusion of \mhiggs severely constrains the model. 
For FN in case of inverted hierarchy we again, as for MN with normal hierarchy, performed an additional fit without including the Higgs mass. As expected from the discussion in \secref{resmin} the pull of the top-quark diminishes and we get $\chi^2_{\rm min} = 0.67$  to be compared to $\chi^2_{\rm min} = 13.3$ in case the Higgs mass is included in the fit. We also fit the model with the 
lower top-quark mass $168.9 \pm 3.5$~GeV, finding $\chi^2_{\rm min} = 0.98$.  We summarize different fits with heavy top-quark, with light top-quark and fits without Higgs mass in \tabref{lighttop}.


\begin{table}[t]
\centering
\begin{tabular}{@{}ccccc@{}}
\toprule
			&	\multicolumn{2}{c}{FS, NH, RGE} &	\multicolumn{2}{c}{FS, IH, RGE}  \\ 
\cmidrule(l){2-3} \cmidrule(l){4-5}
Observable  &  best-fit  &  pull  &  best-fit  &  pull	\\
\midrule
$m_d$ & 0.00290 & -0.0001 & 0.00305 & 0.1255  \\
$m_s$ & 0.05496 & -0.0025 & 0.04337 & -0.7500  \\
$m_b$ & 2.89002 & 0.0003 & 2.88344 & -0.0729  \\
$m_u$ & 0.00127 & -0.0000 & 0.00127 & -0.0062  \\
$m_c$ & 0.61903 & 0.0003 & 0.61551 & -0.0416  \\
$m_t$ & 171.699 & -0.0003 & 171.655 & -0.0150  \\
$\Delta m_{21}^2$ & 7.50$\times 10^{-5}$ & 0.0000 & 7.50$\times 10^{-5}$ & 0.0000  \\
$\Delta m_{31}^2$ & 2.47$\times 10^{-3}$ & -0.0000 & -2.36$\times 10^{-3}$ & -0.0000  \\
$\sin \theta^q_{12}$ & 0.22460 & 0.0000 & 0.22460 & -0.0012  \\
$\sin \theta^q_{23}$ & 0.04200 & 0.0003 & 0.04201 & 0.0114  \\
$\sin \theta^q_{13}$ & 0.00350 & -0.0000 & 0.00350 & -0.0015  \\
$\delta_{\rm CKM}$ & 1.21528 & 0.0001 & 1.21507 & -0.0035  \\
$\sin^2 \theta^l_{12}$ & 0.30000 & 0.0000 & 0.30000 & -0.0000  \\
$\sin^2 \theta^l_{23}$ & 0.41000 & -0.0002 & 0.58972 & -0.0129  \\
$\sin^2 \theta^l_{13}$ & 0.02300 & -0.0003 & 0.02300 & 0.0011  \\
$m_e$ & 0.00049 & 0.0000 & 0.00049 & -0.0025  \\
$m_{\mu}$ & 0.10272 & 0.0001 & 0.10320 & 0.0937  \\
$m_{\tau}$ & 1.74622 & -0.0002 & 1.75356 & 0.0838  \\
\midrule
$\chi^2_{\rm min}$ &       &  $\mathbf{ 6.89\times 10^{-6} }$ &     & {\bf 0.602 } \\
\bottomrule
\end{tabular}
\caption{18 Observables and pulls for model FS (18 free parameters) fitted assuming normal (NH) or inverted (IH) hierarchy. Masses are given in GeV, mass-squared differences in eV$^2$. The value $\tan \beta = 50$ is chosen here, 
with very little difference to other values.}
\label{tab:fsntpulls}
\end{table}

For fits where $\chi^2_{\rm min} \sim 0$ the best-fit values of observables are essentially identical with the experimental values presented in \secref{fittingprocedure}. Hence, we do not tabulate them.
In the non-SUSY fits including RG evolution  we have again the dominating contribution to $\chi^2_{\rm min}$ from $m_t$ with a pull of $\sim -3.25$ (NH) or $\sim -3.5$ (IH) corresponding to $\Delta \chisq_{m_t} \sim 10.6$ (NH) or $\Delta \chisq_{m_t} \sim 12.2$ (IH) followed by the pulls of $\lambda$ and the strange-quark mass $m_s$, 0.65 and 0.45 (NH) or 0.69 and 0.74 (IH), respectively. 
In the SUSY case with normal hierarchy the main contribution to $\chi^2_{\rm min}$ is the strange-quark mass with a pull of 0.37 followed by $\sin^2 \theta_{13}^l$, $m_d$, and $\sin^2 \theta_{23}^l$. 
Fitting the inverted hierarchy, again the pull of the strange-quark mass gives the main contribution to $\chi^2_{\rm min}$ with now the pull being $-0.75$ accounting for nearly the whole value of $\chi^2_{\rm min}$. The best-fit values of observables and their pulls for the non-SUSY and SUSY version of this model are summarized in \tabref{f1ntpulls,fsntpulls}, respectively.

\subsection{Model Predictions}
\label{sec:predictions}
There are observables which have not yet been measured experimentally but are fixed by the fits we performed, so they can be understood as predictions of the models we analyzed. 
For instance, the effective mass \meff 
that governs the lifetime of neutrinoless double beta decay (\nubb), defined as 
\cite{Rodejohann:2011mu,Rodejohann:2012xd},
\beq
\meff = \left| U_{e1}^2 \, m_1 + U_{e2}^2 \, m_2 \, e^{i \alpha} 
+ U_{e3}^2 \, m_3 \, e^{i\beta}\right| \, ,
\ee
is of interest. Here $U$ is the leptonic mixing matrix, $\alpha, \beta$ are Majorana phases and $m_i$ are the masses of light neutrinos. Additional parameters of interest are the leptonic $CP$ violation phase $\delta_{CP}^l$ as relevant for oscillation experiments, and the lightest neutrino mass $m_0$ ($m_0 = m_1$ for NH and $m_0 = m_3$ for IH).  We also present the masses of heavy neutrinos $M_i$, though those are not really testable observables. We will discuss the non-SUSY case as well as the SUSY case. In case of SUSY models we will restrict the discussion to models with $\tan \beta = 50$, since the results of models with other values of $\tan \beta$ are very similar. The numerical values are tabulated in \tabref{predictions}.

\begin{figure}[t]
\centering
\subfloat[]{\label{fig:chivss23-ih}\epsfig{file=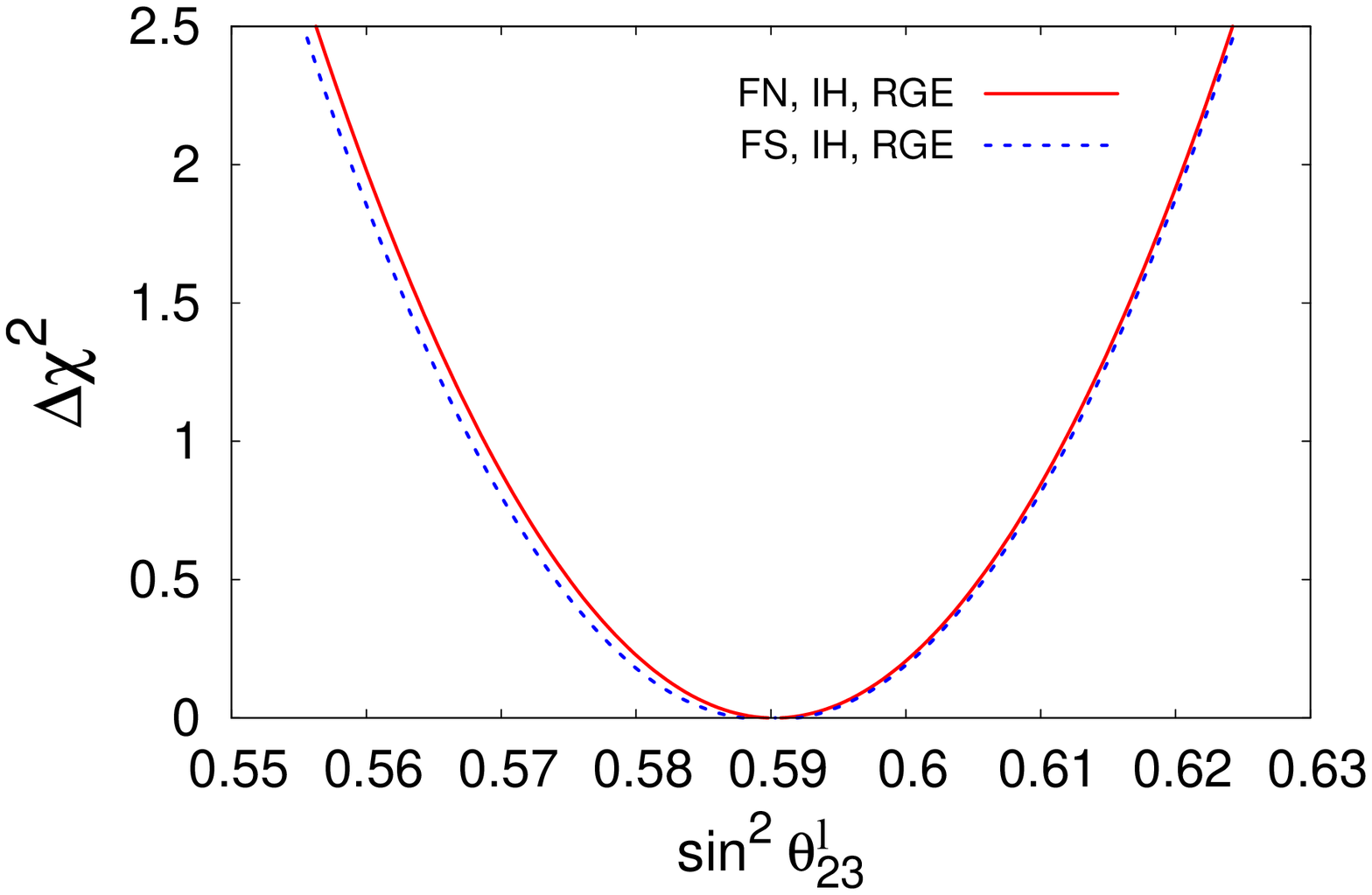,width=7cm,height=6cm}}\hspace{.05cm}
\subfloat[]{\label{fig:chivss23-nh}\epsfig{file=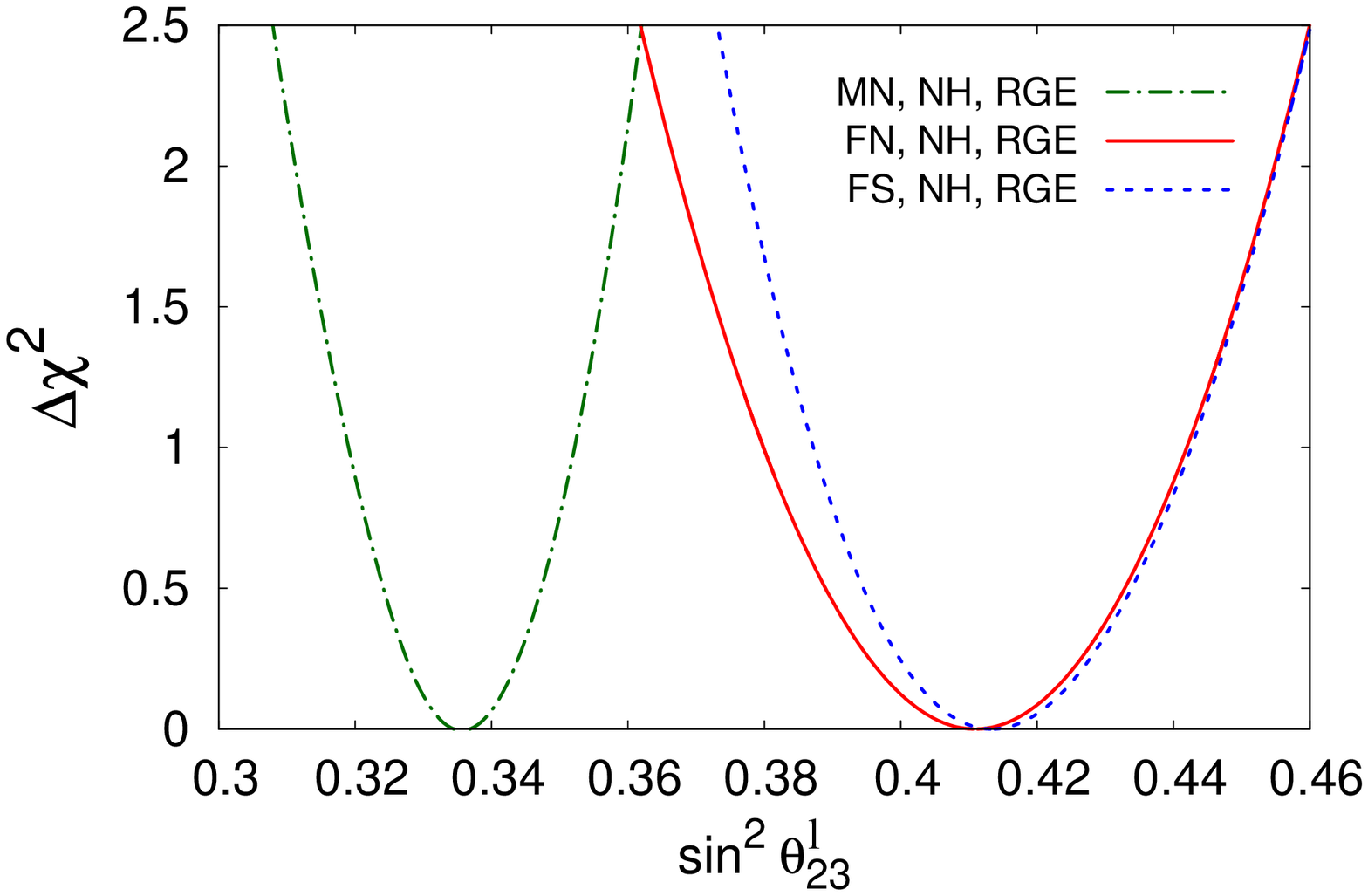,width=8.5cm,height=6cm}}
\caption{$\Delta \chisq(\sin^2 \theta_{23}^l) = \chisq(\sin^2 \theta_{23}^l) - \chisq|_{\rm min}$ is shown  \protect\subref{fig:chivss23-ih} for models FN and FS in case of an inverted neutrino mass hierarchy and
\protect\subref{fig:chivss23-nh} for models MN, FN, and FS in case of a normal hierarchy.}
\label{fig:chivss23}
\end{figure}

There is still the question whether the atmospheric neutrino mixing
angle, $\theta_{23}^l$, deviates from maximal mixing. While the
best-fit points of global neutrino fits usually are away from maximal,
this is typically only a less than $2\,\sigma$ effect (see
\tabref{expvals}). In most of our fits the value of $\theta_{23}^l$ is
fitted essentially at its best-fit point. The notable exception is
model MN in case of a normal hierarchy, where $\sin^2 \theta_{23}^l$
is off by about $2 \, \sigma$, cf.\ Tbl.\ \ref{tab:m1ntresults}. 
We analyze the behaviour of \chisq as a function of $\sin^2 \theta_{23}^l$, 
employing the method described at the beginning of \secref{fittingprocedure}. 
For the still viable models MN, FN, and FS\footnote{MN is not viable in the case of inverted hierarchy, as discussed in \secref{resmin}} we plot $\Delta \chisq(\sin^2 \theta_{23}^l) = \chisq(\sin^2 \theta_{23}^l) - \chi^2_{\rm min}$ in Fig.\ \ref{fig:chivss23}. As can be seen from Fig.~\ref{fig:chivss23} neither model FN nor FS restricts the value of $\theta_{23}^l$ sizably beyond its experimental boundaries, independently of the neutrino mass hierarchy; $\Delta \chisq(\sin^2 \theta_{23}^l)$ simply increases due to the deviation of $\sin^2 \theta_{23}^l$ from the experimental best-fit value. 
Model MN in case of normal hierarchy, however, strongly favors a rather small value, $\sin^2 \theta_{23}^l = 0.335 \pm 0.015$ at $68\,\%$ C.L.\ (corresponding to $\Delta \chisq = 1$), with a much steeper $\Delta \chisq(\sin^2 \theta_{23}^l)$ function. Therefore, if after more precise measurements the value of $\theta_{23}^l$ contracts around its current experimental best-fit value, model MN will be strongly disfavored. 
For other cases, not shown in Fig.\ \ref{fig:chivss23}, especially the fits without RGE, 
the situation is similar to that of models FN and FS as presented in Fig.\ \ref{fig:chivss23}.

No general conclusions can be drawn for leptonic $CP$ phase
$\delta_{CP}^l$ (this may be different if in future analyses the
baryon asymmetry of the Universe as generated via thermal leptogenesis
is also fitted). However, for neutrino masses, $m_i$, $M_i$ and $\meff
$, one can observe that in the normal hierarchy, models FN and FS predict a
higher seesaw scale ($M_3$) than models MN and MS. The
light neutrino masses and the effective mass for neutrinoless double
beta decay are also larger in models FN and FS. 

As can be seen in \tabref{predictions}, the values of essentially all parameters depend strongly on 
whether RG effects are taken into account or not. This shows quite strongly the necessity to 
include them.

\begin{table}[t]
\centering
\begin{tabular}{@{}llccccllll@{}}
\toprule
	  &  	  		&  \meff   	&  $\delta_{CP}^l$  &
          $\sin^2 \theta^l_{23} $ &  $m_0$  & $M_3$  &  $M_2$  &
          $M_1$ & $\chi^2_{\rm min}$ \\
Mod & Comments	&	[meV]	& 	[rad]			&		& [meV]	 & [GeV] & [GeV] & [GeV] &  \\
\midrule
MN & no RGE, NH & 0.35 & 0.7 		& 0.406	& 3.03 & 5.5$\times 10^{12}$ & 7.2$\times 10^{11}$ & 1.5$\times 10^{10}$ & 1.10 \\
MN & RGE, NH & 0.49 & 6.0 	& 0.346	& 2.40 & 3.6$\times 10^{12}$ & 2.0$\times 10^{11}$ & 1.2$\times 10^{11}$ & 23.0 \\
MS & no RGE, NH &  0.38 & 0.27 		& 0.387	& 2.58 & 3.9$\times 10^{12}$ & 7.2$\times 10^{11}$ & 1.6$\times 10^{10}$ & 9.41 \\
MS & RGE, NH & 0.44 & 2.8 	& 0.410	& 6.83 & 1.1$\times 10^{12}$ & 5.7$\times 10^{10}$ & 1.5$\times 10^{10}$ & 3.29 \\
\midrule
FN & no RGE, NH & 4.96 & 1.7 		& 0.410	& 8.8 & 1.9$\times 10^{13}$ & 2.8$\times 10^{12}$ & 2.2$\times 10^{10}$ & 6.6$\times 10^{-5}$ \\
FN & RGE, NH & 2.87 & 5.0 	& 0.410	& 1.54 & 9.9$\times 10^{14}$ & 7.3$\times 10^{13}$ & 1.2$\times 10^{13}$ & 11.2 \\
FS &  no RGE, NH & 0.75 & 0.5 		& 0.410	& 1.16 & 1.5$\times 10^{13}$ & 5.3$\times 10^{11}$ & 5.7$\times 10^{10}$ & 9.0$\times10^{-10}$ \\
FS &  RGE, NH & 0.78 & 5.4 	& 0.410	& 3.17 & 4.2$\times 10^{13}$ & 4.9$\times 10^{11}$ & 4.9$\times 10^{11}$ & 6.9$\times 10^{-6}$ \\
\midrule
FN &  no RGE, IH & 35.37 & 5.4 		& 0.590	& 35.85 & 2.2$\times 10^{13}$ & 4.9$\times 10^{12}$ & 9.2$\times 10^{11}$ & 2.5$\times 10^{-4}$ \\
FN &  RGE, IH & 35.52 & 4.7 & 0.590	& 30.24 & 1.1$\times 10^{13}$ & 3.5$\times 10^{12}$ & 5.5$\times 10^{11}$ & 13.3 \\
FS &  no RGE, IH & 44.21 & 0.3 		& 0.590	& 6.27 & 1.2$\times 10^{13}$ & 4.2$\times 10^{11}$ & 3.5$\times 10^{7}$ & 3.9$\times 10^{-8}$ \\
FS &  RGE, IH & 24.22 & 3.6 & 0.590	& 11.97 & 1.2$\times 10^{13}$ & 3.1$\times 10^{11}$ & 2.0$\times 10^{3}$ & 0.602 \\
\bottomrule
\end{tabular}
\caption{Model predictions for effective \nubb mass \meff, leptonic $CP$ violation $\delta_{CP}^l$, 
atmospheric neutrino mixing parameter $\sin^2 \theta^l_{23}$,  
lightest neutrino mass $m_0$ ($m_0 = m_1$ for NH and $m_0 = m_3$ for IH), and masses of heavy neutrinos $M_i$. For the SUSY models (MS, FS) the predictions with $\tan \beta = 50$ are shown which do not differ significantly from predictions with other values of $\tan \beta$. Models MN and MS have Higgs representations $10_H + \overline{126}_H$, models FN and FS have Higgs representations $10_H + 120_H + \overline{126}_H$. Models not included in this table do not give a good fit.}
\label{tab:predictions}
\end{table}

\section{Conclusions}
\label{sec:sotenconcl}
In general, Grand Unified Theories and in particular models based on \soten offer intriguing 
frameworks to find an answer to the question of the origin of fermion masses and mixings. 
The various Yukawa coupling matrices of the fermions are 
related, and fits of those relations offer tests of the validity of the models. 
In this work we analyzed renormalizable \soten models based on $10_H + \overline{126}_H$, $120_H + \overline{126}_H$, and $10_H + 120_H + \overline{126}_H$ Higgs representations, assuming type I seesaw dominance. We considered non-supersymmetric as well as supersymmetric models with different values of $\tan \beta$. More model-dependent effects of intermediate scales were neglected in our study, possibly 
influencing the results. 

In non-SUSY models there is a connection between the RGE of fermion parameters and the RGE of the Higgs quartic coupling, which is especially important for the top-quark Yukawa coupling. From the Higgs quartic coupling and the Yukawa couplings of fermions at \mgut, the quartic coupling and hence the Higgs mass gets determined at low energies as well. Therefore we included the Higgs mass into our list of observables, 
which has not been done in the literature so far. 
Through their RGE interplay, 
the somewhat low mass of the Higgs leads to a preference for a low top-quark mass, 
as discussed in \secref{resmin} in detail. 

Further, we performed a complete 1-loop RGE from high scale, where the GUT relations for the various Yukawa matrices are valid, to the weak scale, at which the experimental data is available. 
This is a more consistent procedure than doing it the other way around. 
In addition, we treated right-handed neutrinos during RGE correctly and integrated them out one by one at their respective energy scales. 

Finally, we gave the model predictions for several as yet unmeasured observables. These are the effective mass \meff relevant for neutrinoless double beta decay, the leptonic $CP$ violating phase, $\delta_{CP}^l$, the mass of the lightest neutrino, and, for completeness, the masses of the heavy neutrinos.

The results of our analysis are as follows: 
\begin{itemize}
\item 
We showed that it is possible to fit the minimal setups MN and MS\footnote{We remind the reader that model names containing the letter "S" refer to supersymmetric models, while those with "N" refer to non-supersymmetric models.} (both with $10_H + \overline{126}_H$ Higgs representations responsible for fermion mass generation) in the case of the normal neutrino mass hierarchy, while both the non-SUSY (MN) and the SUSY (MS) cases do not work with the inverted hierarchy. 
The alternative minimal model ($120_H + \overline{126}_H$) generates only very large $\chi^2_{\rm min}$-values, 
and is excluded for both possibilities of the neutrino mass hierarchy, 
as well as for the SUSY and non-SUSY cases. 
In contrast, models FN and FS ($10_H + 120_H + \overline{126}_H$) have been shown to be able to reproduce both the normal and inverted hierarchy very well.  
\item For the non-SUSY models (MN and FN) we showed that fitting the Higgs mass leads to severe tensions for the top-quark mass, which is preferred to be more than $3\,\sigma$ smaller than the experimental value. For model FN this is the only observable that cannot be fitted close to its experimental value, while for model MN also $\sin^2 \theta_{23}^l$ deviates significantly, i.e.\ it is about $2 \, \sigma$ smaller than its experimentally measured value. The model sensitively depends on the value of $\sin^2 \theta_{23}^l$. 
\item Regarding the impact of the Higgs mass, we have fitted the models also with the lower top-quark 
mass that is extracted from the total cross section in the top-quark pair production, which has been argued to be more consistent and theoretically more rigorous. As expected, with this lower value of $m_t$ 
the fit improves considerably. 
\item 
An important conclusion is that predictions for 
the unknown parameters \meff, $\delta_{CP}^l$, $\theta_{23}^l$, as well as 
light and heavy neutrino masses, and the value of the \chisq-minimum, depend on whether RGE is included or not. Thus 
we emphasize again the importance of inclusion of RGE when fitting $SO(10)$ models defined at high energy scales.
\end{itemize}


We want to give a few comments on which questions remain open and could be addressed in the future.
First of all, in our CPU-intensive analysis 
we neglected intermediate scales in the breaking scheme of the $SO(10)$ GUT, as well as 
related gauge unification aspects. 
Moreover, the list of models we considered is not exhaustive, so one could analyze further models and compare analyses done with and without RGE. The models we considered either could or could not fit the data, irrespective of considering RGE or not. However, there may well be models where inclusion of RGE makes a difference between considering a model as viable or not. 
Since we restricted our analysis to the type I seesaw case it would be interesting to consider models in which 
either type II seesaw dominates or type I and type II seesaw contributions to neutrino mass are of equal order of magnitude. 
Further, the Yukawa sector is the most unsatisfactory part of gauge theories, as it comes along with a huge number of arbitrary parameters. Aspects such as Yukawa unification or the 
assumption of certain textures in the Yukawa couplings could be 
subject of future studies. This will help in unveiling structures in the Yukawa sector and provide hints to possible fundamental mechanisms governing the Yukawa structure of \soten gauge theories. Finally, 
the baryon asymmetry of the Universe as generated by thermal leptogenesis could be included as 
an additional observable in the fits. 
We leave these modifications and additions to future studies.

\begin{center}
{\bf Acknowledgments}
\end{center}

We thank Thomas Schwetz and Yasutaka Takanishi for helpful discussions, and Rabi Mohapatra for 
illuminating comments.  
This work was supported by the Max Planck Society
through the Strategic Innovation Fund in the project
MANITOP.

\begin{appendix}

\section{Best-Fit Parameters}
\label{sec:bf-params}
%
%
%
%
\subsection{Minimal Models with $10_H + \overline{126}_H$}
Since in these models only the normal neutrino mass hierarchy is viable we give the best-fit parameters only for that case.
\begin{eqnarray} 
& & \underline{\mbox{MN, no RGE:}}  \notag \\
& r & = 68.9624, \, s = 0.370726 + 0.063044 i, \, r_R = 3.014 \times 10^{-16} \text{ GeV}^{-1}  \notag \\ \addlinespace
& H & =
	\left( 
		\begin{smallmatrix} 
			1.22387 \times 10^{-6} & 0 & 0\\
			0 & 5.92428 \times 10^{-5} & 0\\
			0 & 0 & 6.29473 \times 10^{-3}\\
		\end{smallmatrix} 
	\right) \\ \addlinespace
& F & = 
	\left( 
		\begin{smallmatrix} 
			-2.95102 \times 10^{-6}- 3.48291 \times 10^{-6}i & 1.27484 \times 10^{-5}- 7.53714 \times 10^{-8}i & 1.07772 \times 10^{-4}+ 6.02931 \times 10^{-5}i\\
			1.27484 \times 10^{-5}- 7.53714 \times 10^{-8}i & -1.538 \times 10^{-4}+ 6.75236 \times 10^{-5}i & -2.67281 \times 10^{-4}+ 2.48978 \times 10^{-4}i\\
			1.07772 \times 10^{-4}+ 6.02931 \times 10^{-5}i & -2.67281 \times 10^{-4}+ 2.48978 \times 10^{-4}i & -7.38503 \times 10^{-4}- 1.44559 \times 10^{-3}i\\
		\end{smallmatrix} 
	\right) \notag
\end{eqnarray}

\begin{eqnarray}
& & \underline{\mbox{MN, RGE}}  \notag \\
& r & = -63.9043,\, s = 0.409807 - 0.0420522 i, \, r_R = 3.39236 \times 10^{-16} \text{ GeV}^{-1} \notag \\ \addlinespace
 & H & =
        \left(
                \begin{smallmatrix}
                        1.15249 \times 10^{-6} & 0 & 0\\
                        0 & 6.71983 \times 10^{-5} & 0\\
                        0 & 0 & 6.70159 \times 10^{-3}\\
                \end{smallmatrix}
        \right)
 \\ \addlinespace
 & F & =
        \left(
                \begin{smallmatrix}
                        -2.25817 \times 10^{-6}+ 7.40559 \times 10^{-6}i & 1.22057 \times 10^{-5}- 1.39062 \times 10^{-5}i & -1.49653 \times 10^{-4}+ 8.30809 \times 10^{-5}i\\
                        1.22057 \times 10^{-5}- 1.39062 \times 10^{-5}i & -2.06635 \times 10^{-4}- 1.34686 \times 10^{-5}i & 3.76355 \times 10^{-4}+ 2.15049 \times 10^{-4}i\\
                        -1.49653 \times 10^{-4}+ 8.30809 \times 10^{-5}i & 3.76355 \times 10^{-4}+ 2.15049 \times 10^{-4}i & -7.01333 \times 10^{-4}- 7.53673 \times 10^{-4}i\\
                \end{smallmatrix}
        \right)
 \notag
\end{eqnarray}
%
%
%
\begin{eqnarray}
& & \underline{\mbox{MS, } \tan \beta = 50, \text{ no RGE:}}  \notag \\
& r & = 3.21051, \, r_R = 1.46012 \times 10^{-16}  \text{ GeV}^{-1}, \, s = 0.34792 + 0.0110973 i \notag \\ \addlinespace
 & H & = 
	\left( 
		\begin{smallmatrix} 
			1.83386 \times 10^{-5} & 0 & 0\\
			0 & 1.11953 \times 10^{-3} & 0\\
			0 & 0 & 0.1727\\
		\end{smallmatrix} 
	\right)
 \\ \addlinespace
 & F & = 
	\left( 
		\begin{smallmatrix} 
			-5.39376 \times 10^{-5}- 6.82495 \times 10^{-5}i & 2.50369 \times 10^{-4}- 4.11368 \times 10^{-5}i & 2.43574 \times 10^{-3}+ 1.41867 \times 10^{-3}i\\
			2.50369 \times 10^{-4}- 4.11368 \times 10^{-5}i & -2.98376 \times 10^{-3}+ 1.03456 \times 10^{-3}i & -6.4605 \times 10^{-3}+ 6.2383 \times 10^{-3}i\\
			2.43574 \times 10^{-3}+ 1.41867 \times 10^{-3}i & -6.4605 \times 10^{-3}+ 6.2383 \times 10^{-3}i & -0.0106126- 0.0257057i\\
		\end{smallmatrix} 
	\right) \notag
\end{eqnarray}
\begin{eqnarray} 
& & \underline{\mbox{MS, } \tan \beta = 50, \text{ RGE:}}  \notag \\
& r & = 1.87979, \, r_R = -1.09758 \times 10^{-15}  \text{ GeV}^{-1}, \, s = 0.245295 + 0.0935775 i \notag \\ \addlinespace
 & H & =
        \left(
                \begin{smallmatrix}
                        3.61945 \times 10^{-5} & 0 & 0\\
                        0 & 2.77898 \times 10^{-3} & 0\\
                        0 & 0 & 0.627274\\
                \end{smallmatrix}
        \right)
 \\ \addlinespace
 & F & =
        \left(
                \begin{smallmatrix}
                        4.13796 \times 10^{-6}+ 8.08833 \times 10^{-6}i & 5.12296 \times 10^{-4}- 5.07815 \times 10^{-4}i & -1.66274 \times 10^{-3}- 2.47433 \times 10^{-3}i\\
                        5.12296 \times 10^{-4}- 5.07815 \times 10^{-4}i & -8.8214 \times 10^{-3}- 5.30048 \times 10^{-5}i & -0.0170155- 0.019725i\\
                        -1.66274 \times 10^{-3}- 2.47433 \times 10^{-3}i & -0.0170155- 0.019725i & 4.46 \times 10^{-3}- 0.0583555i\\
                \end{smallmatrix}
        \right)
 \notag
\end{eqnarray}
\begin{eqnarray} 
& & \underline{\mbox{MS, } \tan \beta = 38, \text{ no RGE:}}  \notag \\
& r & = 3.43466, \, r_R = 1.66182 \times 10^{-16} \text{ GeV}^{-1}, \, s = 0.347871 + 7.81372 \times 10^{-3} i \notag \\ \addlinespace
 & H & = 
	\left( 
		\begin{smallmatrix} 
			1.969 \times 10^{-5} & 0 & 0\\
			0 & 1.12891 \times 10^{-3} & 0\\
			0 & 0 & 0.16146\\
		\end{smallmatrix} 
	\right)
 \\ \addlinespace
 & F & = 
	\left( 
		\begin{smallmatrix} 
			-5.95751 \times 10^{-5}- 6.59382 \times 10^{-5}i & -2.40742 \times 10^{-4}+ 3.91345 \times 10^{-5}i & -2.42662 \times 10^{-3}- 1.32616 \times 10^{-3}i\\
			-2.40742 \times 10^{-4}+ 3.91345 \times 10^{-5}i & -3.05078 \times 10^{-3}+ 9.77839 \times 10^{-4}i & -6.40902 \times 10^{-3}+ 5.965 \times 10^{-3}i\\
			-2.42662 \times 10^{-3}- 1.32616 \times 10^{-3}i & -6.40902 \times 10^{-3}+ 5.965 \times 10^{-3}i & -9.73364 \times 10^{-3}- 0.0241596i\\
		\end{smallmatrix} 
	\right)
 \notag
\end{eqnarray}
\begin{eqnarray}
& & \underline{\mbox{MS, } \tan \beta = 38, \text{ RGE:}}  \notag \\
& r & = 2.96553, \, r_R = -7.88987 \times 10^{-16} \text{ GeV}^{-1}, \, s = 0.246797 + 0.0669722 i \notag \\ \addlinespace
 & H & =
        \left(
                \begin{smallmatrix}
                        2.24626 \times 10^{-5} & 0 & 0\\
                        0 & 1.63169 \times 10^{-3} & 0\\
                        0 & 0 & 0.312163\\
                \end{smallmatrix}
        \right)
 \\ \addlinespace
 & F & =
        \left(
                \begin{smallmatrix}
                        4.70045 \times 10^{-6}+ 5.21251 \times 10^{-6}i & 3.33325 \times 10^{-4}- 2.99098 \times 10^{-4}i & -7.94881 \times 10^{-4}- 1.5865 \times 10^{-3}i\\
                        3.33325 \times 10^{-4}- 2.99098 \times 10^{-4}i & -5.46865 \times 10^{-3}- 5.53341 \times 10^{-4}i & -8.40926 \times 10^{-3}- 0.0113299i\\
                        -7.94881 \times 10^{-4}- 1.5865 \times 10^{-3}i & -8.40926 \times 10^{-3}- 0.0113299i & -7.01273 \times 10^{-4}- 0.0414157i\\
                \end{smallmatrix}
        \right)
 \notag
\end{eqnarray}
\begin{eqnarray} 
& & \underline{\mbox{MS, } \tan \beta = 10, \text{ no RGE:}}  \notag \\
& r & = 11.9008, \, r_R = 1.66108 \times 10^{-16} \text{ GeV}^{-1}, \, s = 0.352923 + 9.55355 \times 10^{-3} i \notag \\ \addlinespace
 & H & = 
	\left( 
		\begin{smallmatrix} 
			6.75094 \times 10^{-6} & 0 & 0\\
			0 & 3.53643 \times 10^{-4} & 0\\
			0 & 0 & 0.0455808\\
		\end{smallmatrix} 
	\right)
 \\ \addlinespace
 & F & = 
	\left( 
		\begin{smallmatrix} 
			-2.04956 \times 10^{-5}- 1.99197 \times 10^{-5}i & 7.07769 \times 10^{-5}- 8.71429 \times 10^{-6}i & -7.40964 \times 10^{-4}- 3.79993 \times 10^{-4}i\\
			7.07769 \times 10^{-5}- 8.71429 \times 10^{-6}i & -9.71448 \times 10^{-4}+ 2.86615 \times 10^{-4}i & 1.95589 \times 10^{-3}- 1.69133 \times 10^{-3}i\\
			-7.40964 \times 10^{-4}- 3.79993 \times 10^{-4}i & 1.95589 \times 10^{-3}- 1.69133 \times 10^{-3}i & -2.77458 \times 10^{-3}- 7.05915 \times 10^{-3}i\\
		\end{smallmatrix} 
	\right)
 \notag
\end{eqnarray}
\begin{eqnarray} 
& & \underline{\mbox{MS, } \tan \beta = 10, \text{ RGE:}}  \notag \\
& r & = 13.1538, \, r_R = -6.17321 \times 10^{-16} \text{ GeV}^{-1}, \, s = 0.244325 + 0.0495071 i \notag \\ \addlinespace
 & H & =
        \left(
                \begin{smallmatrix}
                        5.12266 \times 10^{-6} & 0 & 0\\
                        0 & 3.60146 \times 10^{-4} & 0\\
                        0 & 0 & 0.0622718\\
                \end{smallmatrix}
        \right)
 \\ \addlinespace
 & F & =
        \left(
                \begin{smallmatrix}
                        1.38859 \times 10^{-6}+ 1.34952 \times 10^{-6}i & -7.94633 \times 10^{-5}+ 6.55427 \times 10^{-5}i & -1.43753 \times 10^{-4}- 3.70193 \times 10^{-4}i\\
                        -7.94633 \times 10^{-5}+ 6.55427 \times 10^{-5}i & -1.25786 \times 10^{-3}- 2.04911 \times 10^{-4}i & 1.64299 \times 10^{-3}+ 2.45787 \times 10^{-3}i\\
                        -1.43753 \times 10^{-4}- 3.70193 \times 10^{-4}i & 1.64299 \times 10^{-3}+ 2.45787 \times 10^{-3}i & -3.6772 \times 10^{-4}- 0.0102974i\\
                \end{smallmatrix}
        \right)
 \notag 
\end{eqnarray}
\subsection{Models with $10_H + \overline{126}_H + 120_H$}
\subsubsection{Normal Neutrino Mass Hierarchy}
\begin{eqnarray} 
& & \underline{\mbox{FN, } \text{no RGE:}}  \notag \\
& r & = 67.1992, \, r_R = -1.54145 \times 10^{-16} \text{ GeV}^{-1}, \, s = -2.0155, \, t_l = 1.09375, \notag \\
& t_u & = -0.973721, \, t_D = -4.11394 \notag \\ \addlinespace
 & H & = 
	\left( 
		\begin{smallmatrix} 
			-1.44349 \times 10^{-3} & 0 & 0\\
			0 & -2.12083 \times 10^{-4} & 0\\
			0 & 0 & 8.38498 \times 10^{-6}\\
		\end{smallmatrix} 
	\right)
	\notag
 \\ \addlinespace
 & F & = 
	\left( 
		\begin{smallmatrix} 
			-3.52616 \times 10^{-3} & -1.87525 \times 10^{-5} & -3.01471 \times 10^{-5}\\
			-1.87525 \times 10^{-5} & -4.43985 \times 10^{-4} & -2.33814 \times 10^{-5}\\
			-3.01471 \times 10^{-5} & -2.33814 \times 10^{-5} & 1.94067 \times 10^{-6}\\
		\end{smallmatrix} 
	\right)
 \\ \addlinespace
 & G & = 
	\left( 
		\begin{smallmatrix} 
			0 & 1.98673 \times 10^{-3} & 1.36719 \times 10^{-4}\\
			-1.98673 \times 10^{-3} & 0 & -1.67619 \times 10^{-5}\\
			-1.36719 \times 10^{-4} & 1.67619 \times 10^{-5} & 0\\
		\end{smallmatrix} 
	\right)
	\notag
\end{eqnarray}
\begin{eqnarray}
& & \underline{\mbox{FN, } \text{RGE:}}  \notag \\
& r & = 63.4279, \, r_R = 1.08547 \times 10^{-18} \text{ GeV}^{-1}, \, s = 0.450438, \, t_l = 3.60171, \notag \\
 & t_u & = -0.0648445, \, t_D = -52.3076 \notag \\ \addlinespace
 & H & =
        \left(
                \begin{smallmatrix}
                        4.1021 \times 10^{-6} & 0 & 0\\
                        0 & 1.29554 \times 10^{-4} & 0\\
                        0 & 0 & 6.78427 \times 10^{-3}\\
                \end{smallmatrix}
        \right)
 \notag \\ \addlinespace
 & F & =
        \left(
                \begin{smallmatrix}
                        -7.62731 \times 10^{-6} & 7.68715 \times 10^{-6} & 3.06531 \times 10^{-5}\\
                        7.68715 \times 10^{-6} & -2.21886 \times 10^{-4} & 5.05238 \times 10^{-4}\\
                        3.06531 \times 10^{-5} & 5.05238 \times 10^{-4} & -7.89186 \times 10^{-4}\\
                \end{smallmatrix}
        \right)
  \\ \addlinespace
 & G & =
        \left(
                \begin{smallmatrix}
                        0 & 3.65588 \times 10^{-5} & -2.26729 \times 10^{-5}\\
                        -3.65588 \times 10^{-5} & 0 & 1.19187 \times 10^{-5}\\
                        2.26729 \times 10^{-5} & -1.19187 \times 10^{-5} & 0\\
                \end{smallmatrix}
        \right)
	\notag
\end{eqnarray}
\begin{eqnarray}
& & \underline{\text{FS, $\tan \beta = 50$, no RGE:}}  \notag \\
& r & = -0.209965, \, r_R = -2.06476 \times 10^{-16} \text{ GeV}^{-1}, \, s = -3.15082, \, t_l = 164.558 \notag \\
&  t_u & = -18.4887, \, t_D = 1.98859 \notag \\ \addlinespace
 & H & = 
	\left( 
		\begin{smallmatrix} 
			-0.0246111 & 0 & 0\\
			0 & 2.11922 & 0\\
			0 & 0 & 1.16561 \times 10^{-3}\\
		\end{smallmatrix} 
	\right) \notag
 \\ \addlinespace
 & F & = 
	\left( 
		\begin{smallmatrix} 
			-6.79587 \times 10^{-3} & 0.0141982 & 9.32493 \times 10^{-6}\\
			0.0141982 & -0.149691 & 4.06145 \times 10^{-3}\\
			9.32493 \times 10^{-6} & 4.06145 \times 10^{-3} & 4.49951 \times 10^{-4}\\
		\end{smallmatrix} 
	\right)
  \\ \addlinespace
 & G & = 
	\left( 
		\begin{smallmatrix} 
			0 & 3.58503 \times 10^{-3} & 4.93814 \times 10^{-5}\\
			-3.58503 \times 10^{-3} & 0 & 1.87896 \times 10^{-4}\\
			-4.93814 \times 10^{-5} & -1.87896 \times 10^{-4} & 0\\
		\end{smallmatrix} 
	\right) \notag
\end{eqnarray}
\begin{eqnarray}
& & \underline{\text{FS, $\tan \beta = 50$, RGE:}}  \notag \\
& r & = -2.26973, \, r_R = -1.40822 \times 10^{-16} \text{ GeV}^{-1}, \, s = 0.528664, \, t_l = -1.31887, \notag \\
 & t_u & = 0.598706, \, t_D = -0.206913 \notag \\ \addlinespace
 & H & =
        \left(
                \begin{smallmatrix}
                        1.49174 \times 10^{-4} & 0 & 0\\
                        0 & 4.89692 \times 10^{-3} & 0\\
                        0 & 0 & 0.327351\\
                \end{smallmatrix}
        \right)
 \notag \\ \addlinespace
 & F & =
        \left(
                \begin{smallmatrix}
                        9.74987 \times 10^{-4} & -3.17774 \times 10^{-3} & -6.38013 \times 10^{-3}\\
                        -3.17774 \times 10^{-3} & -5.84881 \times 10^{-4} & -8.33647 \times 10^{-3}\\
                        -6.38013 \times 10^{-3} & -8.33647 \times 10^{-3} & 0.302701\\
                \end{smallmatrix}
        \right)
  \\ \addlinespace
 & G & =
        \left(
                \begin{smallmatrix}
                        0 & -7.92375 \times 10^{-4} & 0.0323128\\
                        7.92375 \times 10^{-4} & 0 & -0.084999\\
                        -0.0323128 & 0.084999 & 0\\
                \end{smallmatrix}
        \right) \notag
\end{eqnarray}
\subsubsection{Inverted Neutrino Mass Hierarchy}
\begin{eqnarray}
& & \underline{\text{FN, no RGE:}}  \notag \\
& r & = -71.6954, \, r_R = 6.24335 \times 10^{-16} \text{ GeV}^{-1}, \, s = 0.710962, \, t_l = -11.9888 \notag \\ 
& t_u & = -0.049547, \, t_D = 21.5488 \notag \\ \addlinespace
 & H & = 
	\left( 
		\begin{smallmatrix} 
			6.47261 \times 10^{-3} & 0 & 0\\
			0 & 2.64518 \times 10^{-4} & 0\\
			0 & 0 & 4.10329 \times 10^{-5}\\
		\end{smallmatrix} 
	\right)
\notag \\ \addlinespace
 & F & = 
	\left( 
		\begin{smallmatrix} 
			-8.37936 \times 10^{-4} & -8.11918 \times 10^{-4} & 5.65936 \times 10^{-6}\\
			-8.11918 \times 10^{-4} & -2.65901 \times 10^{-4} & 1.47332 \times 10^{-6}\\
			5.65936 \times 10^{-6} & 1.47332 \times 10^{-6} & -5.74189 \times 10^{-5}\\
		\end{smallmatrix} 
	\right)
  \\ \addlinespace
 & G & = 
	\left( 
		\begin{smallmatrix} 
			0 & -9.67226 \times 10^{-8} & -2.30032 \times 10^{-5}\\
			9.67226 \times 10^{-8} & 0 & -2.9926 \times 10^{-5}\\
			2.30032 \times 10^{-5} & 2.9926 \times 10^{-5} & 0\\
		\end{smallmatrix} 
	\right) \notag
\end{eqnarray}
\begin{eqnarray}
& & \underline{\text{FN, RGE:}}  \notag \\
& r & = -65.5547, \, r_R = 1.15340 \times 10^{-16} \text{ GeV}^{-1}, \, s = 0.666694, \, t_l = -9.90739, \notag \\
& t_u & = -0.0535721, \, t_D = 15.6874 \notag \\ \addlinespace
 & H & =
        \left(
                \begin{smallmatrix}
                        4.28045 \times 10^{-5} & 0 & 0\\
                        0 & 2.67413 \times 10^{-4} & 0\\
                        0 & 0 & 6.59444 \times 10^{-3}\\
                \end{smallmatrix}
        \right)
 \notag \\ \addlinespace
 & F & =
        \left(
                \begin{smallmatrix}
                        -6.35416 \times 10^{-5} & 2.97446 \times 10^{-6} & -2.10727 \times 10^{-6}\\
                        2.97446 \times 10^{-6} & -2.90747 \times 10^{-4} & -8.51248 \times 10^{-4}\\
                        -2.10727 \times 10^{-6} & -8.51248 \times 10^{-4} & -6.38279 \times 10^{-4}\\
                \end{smallmatrix}
        \right)
  \\ \addlinespace
 & G & =
        \left(
                \begin{smallmatrix}
                        0 & 3.89049 \times 10^{-5} & 3.20285 \times 10^{-5}\\
                        -3.89049 \times 10^{-5} & 0 & -7.38437 \times 10^{-7}\\
                        -3.20285 \times 10^{-5} & 7.38437 \times 10^{-7} & 0\\
                \end{smallmatrix}
        \right) \notag
\end{eqnarray}
\begin{eqnarray}
& & \underline{\text{FS, $\tan \beta = 50$, no RGE:}}  \notag \\
& r & = 4.05193, \, r_R = 2.62768\times 10^{-16} \text{ GeV}^{-1}, \, s = 0.146437, \, t_l = -1.88674 \notag \\ 
& t_u & = -0.143497, \, t_D = -5.43533 \times 10^{-3} \notag \\ \addlinespace
 & H & = 
	\left( 
		\begin{smallmatrix} 
			-5.64699 \times 10^{-5} & 0 & 0\\
			0 & -0.111019 & 0\\
			0 & 0 & -9.28316 \times 10^{-4}\\
		\end{smallmatrix} 
	\right) \notag
 \\ \addlinespace
 & F & = 
	\left( 
		\begin{smallmatrix} 
			8.11456 \times 10^{-7} & 1.29532 \times 10^{-4} & 3.93343 \times 10^{-5}\\
			1.29532 \times 10^{-4} & -0.159366 & 0.0133342\\
			3.93343 \times 10^{-5} & 0.0133342 & 4.41386 \times 10^{-3}\\
		\end{smallmatrix} 
	\right)
  \\ \addlinespace
 & G & = 
	\left( 
		\begin{smallmatrix} 
			0 & -3.1015 \times 10^{-3} & -7.83212 \times 10^{-4}\\
			3.1015 \times 10^{-3} & 0 & 1.02167 \times 10^{-3}\\
			7.83212 \times 10^{-4} & -1.02167 \times 10^{-3} & 0\\
		\end{smallmatrix} 
	\right) \notag
\end{eqnarray}
\begin{eqnarray}
& & \underline{\text{FS, $\tan \beta = 50$, RGE:}}  \notag \\
& r & = -2.76923, \, r_R = 5.24649 \times 10^{-16} \text{ GeV}^{-1}, \, s = 0.239831, \, t_l = -1.23885 \\
& t_u & = -0.0166744, \, t_D = 1.12799 \times 10^{-6} \notag \\ \addlinespace
 & H & = 
	\left( 
		\begin{smallmatrix} 
			0.327384 & 0 & 0\\
			0 & -5.97675 \times 10^{-7} & 0\\
			0 & 0 & 2.39637 \times 10^{-3}\\
		\end{smallmatrix} 
	\right)
 \notag \\ \addlinespace
 & F & = 
	\left( 
		\begin{smallmatrix} 
			0.309553 & 3.84441 \times 10^{-6} & -0.0201878\\
			3.84441 \times 10^{-6} & -3.51744 \times 10^{-10} & 1.65759 \times 10^{-6}\\
			-0.0201878 & 1.65759 \times 10^{-6} & -6.76014 \times 10^{-3}\\
		\end{smallmatrix} 
	\right)
 \\ \addlinespace
 & G & = 
	\left( 
		\begin{smallmatrix} 
			0 & 1.98713 \times 10^{-3} & 0.0152033\\
			-1.98713 \times 10^{-3} & 0 & -1.47992 \times 10^{-3}\\
			-0.0152033 & 1.47992 \times 10^{-3} & 0\\
		\end{smallmatrix} 
	\right) \notag
\end{eqnarray}
%
%
%
%
%
%
%
%
%
%
%
%
%
%
%
%
%
%
%
%
%
%
%
%
%
%
%
%
%
\section{Beta-Functions for RG Evolution}
\label{sec:betafcts}
To calculate the RGE of observables, RG equations for all parameters of the model under consideration
have to be solved simultaneously. Here we summarize 1-loop RG equations for the SM and the MSSM extended by 
an arbitrary number of right-handed singlet neutrinos. The notation is as in Refs.\ \cite{Antusch:2005gp,Antusch:2002rr}. In particular,
we denote a quantity between the $n$th and $(n+1)$th mass threshold with a superscript $(n)$. 
For further details including 2-loop beta-functions, we refer the reader 
to \cite{Antusch:2005gp,Machacek:1983tz,Machacek:1983fi,Machacek:1984zw,Das:2000uk,Luo:2002ey}.

The beta-functions of the gauge couplings are not affected
by the additional singlets at 1-loop order. They are given by 
\begin{eqnarray}
  16\pi^2\,\beta_{g_A} \equiv 16\pi^2\,\mu \frac{\D g_A}{\D \mu} =b_A \, g_A^3 \;,
\end{eqnarray} 
with $(b_{\SU (3)_\mathrm{C}},b_{\SU (2)_\mathrm{L}},b_{\U (1)_\mathrm{Y}})
=(-7,-\tfrac{19}{6},\tfrac{41}{10})$ in the SM and $(-3,1,\tfrac{33}{5})$ in the MSSM.
For the $\U(1)_\mathrm{Y}$ charge we use GUT normalization.
\subsection{Beta-Functions in the Extended SM}
\label{sec:betafctsSM}
The $\beta$-functions governing RG evolution in the SM extended by singlet neutrinos
are given by \cite{Grzadkowski:1987tf,Antusch:2002rr,Antusch:2005gp} 
{
\allowdisplaybreaks
\begin{subequations}
\begin{eqnarray}
16\pi^2\accentset{(n)}{\beta}_\kappa & = & 
 -\frac{3}{2} (Y_e^\dagger Y_e)^T \:\accentset{(n)}{\kappa}
 -\frac{3}{2}\,\accentset{(n)}{\kappa} \, (Y_e^\dagger Y_e)
 + \frac{1}{2} \RaiseBrace{\bigl(} \accentset{(n)}{Y}^\dagger_D   
   \accentset{(n)}{Y}_D \RaiseBrace{\bigr)}^T \,
  \accentset{(n)}{\kappa}
 +\frac{1}{2}\,\accentset{(n)}{\kappa} \: \RaiseBrace{\bigl(}
 \accentset{(n)}{Y}^\dagger_D\accentset{(n)}{Y}_D\RaiseBrace{\bigr)}
\nonumber \\*
&& {}\vphantom{\frac{1}{2}}
 +2\,\Tr(Y_e^\dagger Y_e)\,\accentset{(n)}{\kappa} 
 +2\, \Tr \RaiseBrace{\bigl(} \accentset{(n)}{Y}^{\dagger}_D 
 \accentset{(n)}{Y}_D\RaiseBrace{\bigr)}\,\accentset{(n)}{\kappa} 
 +6\,\Tr(Y_u^\dagger Y_u)\,\accentset{(n)}{\kappa} 
  \nonumber \\*
 && {} \vphantom{\frac{1}{2}}
 +6\,\Tr(Y_d^\dagger Y_d)\,\accentset{(n)}{\kappa}
- 3 g_2^2\: \accentset{(n)}{\kappa}
 +\lambda\accentset{(n)}{\kappa}
 \;,\\
16\pi^2 \accentset{(n)}{\beta}_{M} &=&\vphantom{\frac{1}{2}}
 \RaiseBrace{\bigl(}\accentset{(n)}{Y}_D   
   \accentset{(n)}{Y}^\dagger_D \RaiseBrace{\bigr)}\, \accentset{(n)}{M} 
   + \accentset{(n)}{M}\,\RaiseBrace{\bigl(}\accentset{(n)}{Y}_D   
   \accentset{(n)}{Y}^\dagger_D \RaiseBrace{\bigr)}^T \;,\\
   16\pi^2 \accentset{(n)}{\beta}_{Y_D}
 & = &
 \accentset{(n)}{Y}_D \left\{ 
        \frac{3}{2} \RaiseBrace{\bigl(}
        \accentset{(n)}{Y}^\dagger_D\accentset{(n)}{Y}_D\RaiseBrace{\bigr)}
        - \frac{3}{2}(Y_e^\dagger Y_e)
+ \Tr \RaiseBrace{\bigl(}\accentset{(n)}{Y}^{\dagger}_D  
\accentset{(n)}{Y}_D\RaiseBrace{\bigr)} +\Tr (Y_e^\dagger Y_e) \right.
\nonumber \\
&&\hphantom{\accentset{(n)}{Y}_D \left[ \right.} \left. 
{}+ 3\,\Tr(Y_u^\dagger Y_u)+3\,\Tr(Y_d^\dagger Y_d)
-\frac{9}{20} g_1^2 -\frac{9}{4} g_2^2 \right\} ,
\\
        16\pi^2 \, \accentset{(n)}{\beta}_{Y_e}
        & = &
        Y_e
        \left\{ 
        \vphantom{\Tr\left[Y_e^\dagger Y_e
                +z_\nu^{(1)}\,\accentset{(n)}{Y}_D^\dagger \accentset{(n)}{Y}_D
                +3z_d^{(1)}\,Y_d^\dagger Y_d\
                +3z_u^{(1)}\,Y_u^\dagger Y_u\right]}
     \frac{3}{2} Y_e^\dagger Y_e 
         -\frac{3}{2}\, \accentset{(n)}{Y}_D^\dagger \accentset{(n)}{Y}_D 
         -       \frac{9}{4} g_1^2 - \frac{9}{4} g_2^2
         \right.\nonumber\\*
        & &\hphantom{Y_e\left\{ \right.}
         \left.{}+
         \Tr\left[Y_e^\dagger Y_e
                +\accentset{(n)}{Y}_D^\dagger \accentset{(n)}{Y}_D
                +3\,Y_d^\dagger Y_d\
                +3\,Y_u^\dagger Y_u\right]
        \right\} ,\\
        16\pi^2 \, \accentset{(n)}{\beta}_{Y_d}
        & = &
        Y_d
        \left\{ 
     \frac{3}{2} Y_d^\dagger Y_d 
         -\frac{3}{2}\, Y_u^\dagger Y_u          
         - \frac{1}{4} g_1^2 - \frac{9}{4} g_2^2 - 8\,g_3^2
         \right.\nonumber\\*
        & &\hphantom{Y_e\left\{ \right.}
         \left.{}+ \Tr\left[Y_e^\dagger Y_e
                +\accentset{(n)}{Y}_D^\dagger \accentset{(n)}{Y}_D
                +3\,Y_d^\dagger Y_d
                +3\,Y_u^\dagger Y_u\right]
        \right\} , 
        \\
        16\pi^2 \, \accentset{(n)}{\beta}_{Y_u}
        & = &
        Y_u
        \left\{ 
     \frac{3}{2} Y_u^\dagger Y_u 
         - \frac{3}{2}\, Y_d^\dagger Y_d  
         - \frac{17}{20} g_1^2 - \frac{9}{4} g_2^2 - 8\,g_3^2
         \right.\nonumber\\*
        & &\hphantom{Y_e\left\{ \right.}
         \left.{}+ \Tr\left[Y_e^\dagger Y_e
                +\accentset{(n)}{Y}_D^\dagger \accentset{(n)}{Y}_D
                +3\,Y_d^\dagger Y_d
                +3\,Y_u^\dagger Y_u\right]
        \right\} ,
\\
 16\pi^2\,\accentset{(n)}{\beta}_{\lambda}
 &=& 
 6\,\lambda^2 
 -3\,\lambda\,\left(3g_2^2+\frac{3}{5} g_1^2\right)
 +3\,g_2^4
 +\frac{3}{2}\,\left(\frac{3}{5} g_1^2+g_2^2\right)^2
 \nonumber
 \\
 & &{}
 +4\,\lambda\,
 \Tr\left[
       Y_e^\dagger Y_e
       +\accentset{(n)}{Y}_D^\dagger\accentset{(n)}{Y}_D
       +3\,Y_d^\dagger Y_d
       +3\,Y_u^\dagger Y_u
 \right]
 \\*
 & &{}
 -8\,\Tr\left[
  Y_e^\dagger Y_e\,Y_e^\dagger Y_e
  +    \accentset{(n)}{Y}_D^\dagger\accentset{(n)}{Y}_D
               \,\accentset{(n)}{Y}_D^\dagger\accentset{(n)}{Y}_D
       +3\,Y_d^\dagger Y_d\,Y_d^\dagger Y_d
       +3\,Y_u^\dagger Y_u\,Y_u^\dagger Y_u
 \right].\nonumber
\end{eqnarray}
\end{subequations}
}
Note that our convention that the Higgs self-interaction term in the
Lagrangian is $-\frac{\lambda}{4} (\phi^\dagger \phi)^2$.
\subsection{Beta-Functions in the Extended MSSM}
\label{sec:betafctsMSSM}
The 1-loop beta-functions of the MSSM extended by heavy singlet neutrinos are given 
by \cite{Grzadkowski:1987wr,Antusch:2002rr,Antusch:2005gp}
\begin{subequations}
\begin{eqnarray}
16 \pi^2 \,  \accentset{(n)}{\beta}_\kappa & = & \vphantom{\frac{1}{2}}
 (Y_e^\dagger Y_e)^T \: \accentset{(n)}{\kappa}
 + \accentset{(n)}{\kappa} \, (Y_e^\dagger Y_e)
 + \RaiseBrace{\bigl(} \accentset{(n)}{Y}^\dagger_D   
   \accentset{(n)}{Y}_D \RaiseBrace{\bigr)}^T\,\accentset{(n)}{\kappa}
 + \accentset{(n)}{\kappa} \: \RaiseBrace{\bigl(}
 \accentset{(n)}{Y}^\dagger_D\accentset{(n)}{Y}_D\RaiseBrace{\bigr)}
\nonumber \\
&& {} + 2 \Tr \RaiseBrace{\bigl(} \accentset{(n)}{Y}^{\dagger}_D 
 \accentset{(n)}{Y}_D\RaiseBrace{\bigr)}\,\accentset{(n)}{\kappa}
 +6\Tr( Y_u^\dagger Y_u)\,\accentset{(n)}{\kappa} 
 -\frac{6}{5} g_1^2 \:\accentset{(n)}{\kappa}- 6 g_2^2 \:
 \accentset{(n)}{\kappa}
\;,
\\
\label{eq:BetaMintheMSSM} 16 \pi^2 \,  \accentset{(n)}{\beta}_{M} &=& \vphantom{\frac{1}{2}}
   2\,\RaiseBrace{\bigl(}\accentset{(n)}{Y}_D
   \accentset{(n)}{Y}^\dagger_D \RaiseBrace{\bigr)}\, \accentset{(n)}{M} 
   + 2\,\accentset{(n)}{M}\,\RaiseBrace{\bigl(}\accentset{(n)}{Y}_D   
   \accentset{(n)}{Y}^\dagger_D \RaiseBrace{\bigr)}^T ,
\\
   16 \pi^2 \,  \accentset{(n)}{\beta}_{Y_D}
 &=&
 \accentset{(n)}{Y}_D
 \left\{ 3 \accentset{(n)}{Y}^\dagger_D
\accentset{(n)}{Y}_D + Y_e^\dagger Y_e
+ \Tr \RaiseBrace{\bigl(}\accentset{(n)}{Y}^{\dagger}_D  
\accentset{(n)}{Y}_D\RaiseBrace{\bigr)} +3\Tr (Y_u^\dagger Y_u)
- \frac{3}{5} g_1^2 - 3 g_2^2 \right\} ,
 \\
16 \pi^2 \,  \accentset{(n)}{\beta}_{Y_d} 
 & = & Y_d\left\{
 3Y_d^\dagger Y_d 
 + Y_u^\dagger Y_u 
 + 3\Tr(Y_d^\dagger Y_d) 
 + \Tr(Y_e^\dagger Y_e)
 - \frac{7}{15}g_1^2
 - 3g_2^2 
 - \frac{16}{3}g_3^2
 \right\},
 \nonumber\\
 \\
16 \pi^2 \,  \accentset{(n)}{\beta}_{Y_u} 
 & = & Y_u\left\{
 Y_d^\dagger Y_d 
 + 3 Y_u^\dagger Y_u 
 + \Tr( \accentset{(n)}{Y}_D ^\dagger  \accentset{(n)}{Y}_D ) 
 + 3\Tr(Y_u^\dagger Y_u)
 - \frac{13}{15}g_1^2 
 - 3g_2^2 
 - \frac{16}{3}g_3^2 
 \right\} ,
 \nonumber\\
 \\
16 \pi^2 \, \accentset{(n)}{\beta}_{Y_e} 
 & = & Y_e\left\{
 3Y_e^\dagger Y_e 
 +  \accentset{(n)}{Y}_D ^\dagger  \accentset{(n)}{Y}_D  
 + 3\Tr(Y_d^\dagger Y_d) 
 + \Tr(Y_e^\dagger Y_e)
 - \frac{9}{5}g_1^2 
 - 3g_2^2 
 \right\}.
\end{eqnarray}
\end{subequations}

\thispagestyle{empty}

\end{appendix}

\bibliographystyle{utcaps}
\bibliography{soten}


\end{document}